\documentclass[apj,iop]{emulateapj}

\usepackage{graphics}
\usepackage{natbib}
\usepackage{color}

\newcommand{\snia}{SN~Ia}
\newcommand{\sneia}{SNe~Ia}
\newcommand{\sne}{SNe}
\newcommand{\snf}{SNfactory}

\newcommand{\zpeg}{{\tt ZPEG}}
\newcommand{\nifs}{$^{56}$Ni}
\newcommand{\chisq}{$\chi^2$}
\newcommand{\galex}{{\em GALEX}}

\newcommand{\saltmassmagstep}{$0.085\pm0.028$~mag}

\newcommand{\saltmassmagslope}{$-0.043\pm0.014$~mag/dex}

\newcommand{\baileymassmagstep}{$0.050\pm0.028$~mag}
\newcommand{\saltsjbmassmagstep}{$0.066\pm0.030$~mag}
\newcommand{\baileyssfrmagstep}{$-0.052\pm0.028$~mag}
\newcommand{\saltsjbssfrmagstep}{$-0.038\pm0.030$~mag}
\newcommand{\baileymassmagslope}{$-0.016\pm0.014$~mag/dex}
\newcommand{\saltsjbmassmagslope}{$-0.042\pm0.016$~mag/dex}
\newcommand{\baileyssfrmagslope}{$0.047\pm0.015$~mag/dex}
\newcommand{\saltsjbssfrmagslope}{$0.021\pm0.017$~mag/dex}
\newcommand{\allmassmagstep}{$0.077\pm0.014$~mag}
\newcommand{\allwidemassmagstep}{$0.086\pm0.016$~mag}
\newcommand{\allwidemassgrad}{$0.14\pm0.03$~mag/dex}
\newcommand{\snfpromptfrac}{$67\pm6$\%}

\begin{document}

\title{Host Galaxy Properties and Hubble Residuals of 
Type Ia Supernovae from the Nearby Supernova Factory}

\author
{
    M.~Childress,\altaffilmark{1,2,3,4}
    G.~Aldering,\altaffilmark{1}
    P.~Antilogus,\altaffilmark{5}
    C.~Aragon,\altaffilmark{1}
    S.~Bailey,\altaffilmark{1}
    C.~Baltay,\altaffilmark{6}
    S.~Bongard,\altaffilmark{5}
    C.~Buton,\altaffilmark{7}
    A.~Canto,\altaffilmark{5}
    F.~Cellier-Holzem,\altaffilmark{5}
    N.~Chotard,\altaffilmark{8}
    Y.~Copin,\altaffilmark{8}
    H.~K. Fakhouri,\altaffilmark{1,2}
    E.~Gangler,\altaffilmark{8}
    J.~Guy,\altaffilmark{5} 
    E.~Y. Hsiao,\altaffilmark{1}
    M.~Kerschhaggl,\altaffilmark{7}
    A.~G.~Kim,\altaffilmark{1}
    M.~Kowalski,\altaffilmark{7}
    S.~Loken,\altaffilmark{1,*}
    P.~Nugent,\altaffilmark{9}
    K.~Paech,\altaffilmark{7}
    R.~Pain,\altaffilmark{5}
    E.~Pecontal,\altaffilmark{10}
    R.~Pereira,\altaffilmark{8}
    S.~Perlmutter,\altaffilmark{1,2}
    D.~Rabinowitz,\altaffilmark{6}
    M.~Rigault,\altaffilmark{8}  
    K.~Runge,\altaffilmark{1}
    R.~Scalzo,\altaffilmark{3}
    G.~Smadja,\altaffilmark{8}
    C.~Tao,\altaffilmark{11,12}
    R.~C. Thomas,\altaffilmark{9}
    B.~A.~Weaver,\altaffilmark{13}
    C.~Wu\altaffilmark{5,14}
}

\altaffiltext{1}
{
    Physics Division, Lawrence Berkeley National Laboratory, 
    1 Cyclotron Road, Berkeley, CA, 94720.
}
\altaffiltext{2}
{
    Department of Physics, University of California Berkeley,
    366 LeConte Hall MC 7300, Berkeley, CA, 94720-7300.
}
\altaffiltext{3}
{
    Research School of Astronomy and Astrophysics,
    The Australian National University,
    Mount Stromlo Observatory,
    Cotter Road, Weston Creek ACT 2611 Australia.
}
\altaffiltext{4}
{
    ARC Centre of Excellence for All-sky Astrophysics (CAASTRO).
}
\altaffiltext{5}
{
    Laboratoire de Physique Nucl\'eaire et des Hautes \'Energies,
    Universit\'e Pierre et Marie Curie Paris 6, Universit\'e Paris Diderot Paris 7, CNRS-IN2P3, 
    4 place Jussieu, 75252 Paris Cedex 05, France.
}
\altaffiltext{6}
{
    Department of Physics, Yale University, 
    New Haven, CT, 06250-8121.
}
\altaffiltext{7}
{
    Physikalisches Institut, Universit\"at Bonn,
    Nu\ss allee 12, 53115 Bonn, Germany.
}
\altaffiltext{8}
{
    Universit\'e de Lyon, F-69622, Lyon, France ; Universit\'e de Lyon 1, Villeurbanne ; 
    CNRS/IN2P3, Institut de Physique Nucl\'eaire de Lyon.
}
\altaffiltext{9}
{
    Computational Cosmology Center, Computational Research Division, Lawrence Berkeley National Laboratory, 
    1 Cyclotron Road MS 50B-4206, Berkeley, CA, 94720.
}
\altaffiltext{10}
{
    Centre de Recherche Astronomique de Lyon, Universit\'e Lyon 1,
    9 Avenue Charles Andr\'e, 69561 Saint Genis Laval Cedex, France.
}
\altaffiltext{11}
{
    Centre de Physique des Particules de Marseille , Aix-Marseille Universit\'e , CNRS/IN2P3, 163, avenue de Luminy - Case 902 - 13288 Marseille Cedex 09, France.
}
\altaffiltext{12}
{
    Tsinghua Center for Astrophysics, Tsinghua University, Beijing 100084, China.
}
\altaffiltext{13}
{
    Center for Cosmology and Particle Physics,
    New York University,
    4 Washington Place, New York, NY 10003, USA.
}
\altaffiltext{14}
{
    National Astronomical Observatories, Chinese Academy of Sciences, Beijing 100012, China.
}
\altaffiltext{*}
{
    Deceased.
}

%%%%%%%%%%%%%%%%%%%%%%%%%%%%%%%%%%%%%%%%%%%%%%%%%%%%%%%%%%%%%%%%%%%%%%%%%%%%%
\begin{abstract}
% NEW SNF DATA
We examine the relationship between Type Ia Supernova (\snia) Hubble residuals and the properties of their host galaxies using a sample of 115 \sneia\ from the Nearby Supernova Factory (\snf). 
We use host galaxy stellar masses and specific star-formation rates fitted from photometry for all hosts, as well as gas-phase metallicities for a subset of 69 star-forming (non-AGN) hosts, to show that the SN Ia Hubble residuals correlate with each of these host properties.
% NEW RESULT IN THIS PAPER - color-Z trend!!
With these data we find new evidence for a correlation between \snia\ intrinsic color and host metallicity.
% COMBINED DATA SET RESULTS
When we combine our data with those of other published SN Ia surveys, we find the difference between mean SN Ia brightnesses in low and high mass hosts is \allmassmagstep.
% STRUCTURE
When viewed in narrow (0.2~dex) bins of host stellar mass, the data reveal apparent plateaus of Hubble residuals at high and low host masses with a rapid transition over a short mass range ($9.8 \leq \log(M_*/M_\odot) \leq 10.4$).
% ORIGIN OF TREND - more than just Z along galaxy mass sequence!
Although metallicity has been a favored interpretation for the origin of the Hubble residual trend with host mass, we illustrate how dust in star-forming galaxies and mean \snia\ progenitor age both evolve along the galaxy mass sequence, thereby presenting equally viable explanations for some or all of the observed \snia\ host bias.
\end{abstract}

\keywords{supernovae: general --- cosmology: dark energy}

%%%%%%%%%%%%%%%%%%%%%%%%%%%%%%%%%%%%%%%%%%%%%%%%%%%%%%%%%%%%%%%%%%%%%%%%%%%%%
\section{Introduction}
% generic intro to why we care about sneia
% focus on redshift evolution and progenitor bias!
Type Ia Supernovae (\sneia) are excellent tools for measuring cosmological distances because of their brightness ($M_B \sim -19.5$) and low intrinsic luminosity dispersion ($\sim 0.35$~mag). \sneia\ exhibit even tighter dispersion ($\sim 0.15-0.18$~mag) after application of empirical correction for the shape of their light curves \citep{phillips93, hamuy95, hamuy96, riess95, perlmutter97} and their color \citep{riess96, tripp98}. These qualities make them invaluable tools for measuring the expansion history of the Universe \citep{riess98, perlmutter99}, providing the tightest possible constraints on the matter and dark energy densities as well as the dark energy equation of state parameter $w$ \citep[e.g.][]{riess98, perlmutter99, garnavich98, knop03, riess04, riess07, sullivan11a, suzuki12}. 

Despite the successful empirical utility of \sneia, a thorough census of their stellar progenitors has yet to be established. However, some recent studies have begun to shed light on the likely progenitors of a few individual \sneia. Pre-explosion {\em Hubble Space Telescope (HST)} imaging \citep{li11fe} of the very nearby \snia\ SN~2011fe in M101 \citep{nugent11} ruled out all red giant companions to the exploding star, while the first photometric measurement of SN~2011fe only 4 hours after explosion \citep{bloom12} confidently constrained the exploding star to be a white dwarf and forbade most main sequence companions. {\em HST} imaging of a \snia\ remnant in the Large Magellanic Cloud \citep{schaefer12} ruled out any possible non-degenerate companion to the exploding star \citep[i.e. the single degenerate, SD, scenario][]{whelan73}, thereby strongly favoring a progenitor system comprising two white dwarfs \citep[the double degenerate, DD, scenario][]{iben84}. Conversely, some \sneia\ such as SN~2002ic \citep{mwv04, hamuy03}, SN~2005gj \citep{aldering06, prieto07}, or recently PTF11kx \citep{dilday12} have shown clear signs of interaction with some circum-stellar material, strongly favoring a SD progenitor scenario for these SNe.

Indeed while these individual cases are very exciting and mark a significant step forward in the study of \snia\ progenitors, it is important to build a statistically large sample of \snia\ progenitor detections in order to study the influence of properties such as stellar age and metallicity on the luminosities and light curve shapes of the resultant \sneia. Thus while interesting these individual cases do not yet present a complete picture of \snia\ progenitors, and concerns remain that the lower metallicity and younger ages of high-redshift \snia\ progenitors could potentially alter their luminosities in a way that could bias cosmological parameters.

%------------------
% current hubble/hosts trends
Several recent studies \citep{kelly10,sullivan10,lampeitl10,gupta11,johansson12} observed that the corrected brightnesses of \sneia\ correlated with the stellar masses of their host galaxies, such that \sneia\ in high-mass host galaxies are brighter than \sneia\ in low-mass galaxies after application of the usual stretch- and color-based brightness correction techniques. \citet{gallagher05} found tentative evidence for a similar trend with respect to host galaxy gas-phase metallicity in star-forming hosts which was recently confirmed with higher significance in the SDSS-SN sample by \citet{konishi11} and \citet{dandrea11} \citep[though see][who find tentative evidence for an opposite trend in passive hosts]{gallagher08}, and for photometrically-inferred metallicities by \citet{hayden12}. The origin of this trend is of paramount concern for \snia\ cosmologists. Galaxy mass correlates with metallicity \citep{trem04} and stellar age \citep{gallazzi05}, posing the possibility that this trend could be driven by \snia\ progenitor metallicity or age. The average stellar ages and metallicities of galaxies evolve with redshift, implying that the average corrected \snia\ brightnesses at higher redshift will be fainter than in the local universe if the observed host bias is indeed driven by progenitor age or metallicity. Though this trend is not strong enough to negate the evidence for cosmic acceleration, it could potentially bias estimation of cosmological parameters, especially the dark energy equation of state parameter $w$ and its evolution with redshift. 

%------------------
% what to do about it!
Coincident with the concerns about biased cosmological results is the desire to find a means to correct for this observed \snia\ host bias. Some authors \citep[e.g.][]{sullivan10, lampeitl10} have suggested using host galaxy mass as a third \snia\ brightness correction parameter (after stretch and color). This tactic might ameliorate any progenitor-driven luminosity effects but carries some important complexities to track, such as galaxy stellar population systematics and evolution of galaxy scaling relations with redshift (see Section~\ref{sec:host_corrections} for details). We show later that these considerations are likely to be subdominant in the current \snia\ error budget, and are small compared to the magnitude of the bias the host term is intended to correct, but will add to the complexity of tracking systematic errors in \snia\ cosmological analysis. This further motivates the search for the source of the observed host bias in order to potentially identify a correction technique that minimizes the associated host galaxy systematics. 

%------------------
% purpose of this paper!
In this work we use \snia\ light curve and host galaxy data from the Nearby Supernova Factory \citep[\snf\ --][]{ald02} to add to the body of observational data supporting the existence of a host bias. We then examine possible origins of the trend and prospects for using host-based \snia\ luminosity corrections.
% new summary of papers
This paper is part of planned series of papers examining the host galaxies of \sneia\ from the \snf. The main companion to this paper is Childress et al. 2013a (hereafter Paper~I) which presents the main host galaxy data and derivation of galaxy physical properties (mass, star-formation rate, metallicity) from those data. Paper~III (Childress et al. 2013c, in prep.) explores the implications of \snia\ intrinsic color variability and reddening by host galaxy dust using Monte-Carlo simulation. Finally, Paper~IV (Childress et al. 2013d, in prep.) fits the distribution of \snia\ host galaxy masses by coupling \snia\ delay time distributions to known properties of galaxies in the local Universe.

%------------------
% introduction of paper contents
This paper is organized as follows. In Section~\ref{sec:data} we summarize the \snia\ light curve and host galaxy data used in this analysis. We examine trends of stretch- and color-corrected \snia\ Hubble residuals with host galaxy properties in Section~\ref{sec:salt_trends}, as well as Hubble residuals derived using the spectroscopic standardization technique of \citet{bailey09}. We discuss our results in the context of multiple similar studies and calculate the structure of the host mass Hubble residual trend in Section~\ref{sec:combined_data_sets}. We present several physical processes which may drive this trend in Section~\ref{sec:host_bias_origin}. In Section~\ref{sec:host_corrections} we discuss the implications for \snia\ cosmology analyses if the observed host bias is not corrected, as well as the systematic errors introduced by the application of host-based \snia\ luminosity corrections. Finally, we present our conclusions in Section~\ref{sec:conclusions}. 

% host data and cosmology summary
Host galaxy physical parameters are derived assuming a standard $\Lambda$CDM cosmology with $\Omega_\Lambda = 0.7$, $\Omega_M = 0.3$, $H_0=70$~km~s$^{-1}$~Mpc$^{-1}$ (i.e. $h_{70}=1$), with stellar masses and star-formation rates computed using a \citet{chab03} initial mass function (IMF), as detailed in Paper~I. Cosmological parameters for the \snf\ \snia\ sample are left intentionally blinded to preserve the impartiality of forthcoming \snf\ cosmology analyses.

%%%%%%%%%%%%%%%%%%%%%%%%%%%%%%%%%%%%%%%%%%%%%%%%%%%%%%%%%%%%%%%%%%%%%%%%%%%%%
\section{SN and Host Galaxy Data}
\label{sec:data}
\subsection{\snia\ Data}
% SN sample description
The \sneia\ in this sample were discovered or followed by the \snf\ between 2004 to 2010. Of the 119 fully processed \sneia\ with good light curves in the \snf\ sample, 98 were discovered in the \snf\ search conducted from 2004 to 2008 using the QUEST-II CCD camera \citep{baltay07} on the Samuel Oschin 1.2m Schmidt telescope on Mount Palomar, California. The \snf\ search found many SNe of all types in the low redshift universe, focusing on finding \sneia\ in the redshift range $0.03 < z < 0.08$. Additional \sneia\ presented here were discovered and publicly announced by other nearby searches such as the Lick Observatory Supernova Search \citep{loss} or the Palomar Transient Factory \citep{ptf09}.

% SNIFS data description
The \snia\ light curve data used in this analysis are derived from flux-calibrated spectral time series obtained by \snf\ with the SuperNova Integral Field Spectrograph \citep[SNIFS,][]{lantz04} on the University of Hawaii 2.2m telescope on Mauna Kea. SNIFS is designed for observations of point sources on a diffuse or structured background, and simultaneous observation of field stars in multiple filters allows for spectrophotometric observations through thin clouds. Reduction of SNIFS data is described in \citet{aldering06} and updated in \citet{scalzo10}, \citet{bongard11}, and \citet{buton12}.

% LC description
Photometry was synthesized from SNIFS spectrophotometry at all epochs using non-overlapping boxcar filter functions whose wavelength ranges correspond roughly to Johnson $B$, $V$, and $R$. These light curves were synthesized in the observer frame then fitted using version~2.2 of the SALT2 light curve fitter \citep{guy07,guy10} to derive the peak magnitude in $B$-band ($m_B$) as well as the light curve width ($x_1$) and color ($c$) parameters. Example light curves for interesting \sneia\ from the \snf\ data set can be found in \citet{scalzo10, scalzo12}. Details on our light curve synthesis and fitting techniques can be found in Pereira et al. (2013) as applied to SN~2011fe.

% Hubble diagram, residuals, and sample description
Stretch- and color-corrected Hubble residuals and their errors were calculated in a blinded cosmological analysis similar to that presented in \citet{bailey09}. The sample of \sneia\ examined here includes the sample of \citet{chotard11} but is slightly expanded to include some additional \sneia. For all SNe the structured host galaxies have been subtracted from SNIFS data using the method of \citet{bongard11}. Full details on the cosmological fitting for this sample will be presented in Aldering et al. (2013, in prep.).

% Bailey ratio data!
Later aspects of this analysis (Section~\ref{sec:bailey_trends}) will utilize Hubble residuals derived from \snia\ brightnesses corrected using the spectroscopic standardization technique developed by \citet{bailey09}. In that work we showed that the spectral flux ratio $\mathcal{R}_{642/443}$, defined as the ratio of the \snia\ flux at 642~nm divided by the flux at 443~nm (with 2000~km/s binning), correlated very strongly with the raw \snia\ Hubble residuals (i.e. without stretch and color corrections). We repeat the calculation of $\mathcal{R}_{642/443}$ for the subset of this \snia\ sample with observations within 2.5 days of maximum light. We then calculate a (blinded) luminosity correction factor $\gamma$ and apply the luminosity correction factor $\gamma\mathcal{R}_{642/443}$ to the observed peak brightnesses of our \sneia.

%------------------------------------
\subsection{Host Galaxy Data}
Host galaxy stellar masses, star formation rates (SFRs) and gas-phase metallicities are derived in Paper~I. We here briefly summarize the methods presented there and point to that work and references therein for further details.

%------------------
% photometry
Host galaxy stellar masses and star formation rates (SFRs) were derived by comparing observed multi-band photometry to stellar population synthesis (SPS) models. Photometric data for \snf\ \snia\ host galaxies was gathered from public sources as well as our own targeted observations. Optical photometry was collected from the Sloan Digital Sky Survey \citep[SDSS][]{york00} Eighth Data Release \citep[DR8,][]{aihara11}. NIR images from 2MASS \citep{twomass} were obtained at the NASA/IPAC Infrared Science Archive (IRSA\footnote{http://irsa.ipac.caltech.edu}), with NIR data for some \snia\ hosts from UKIDSS \citep{ukidss}. UV data were obtained from the GALEX online data archive at MAST\footnote{http://galex.stsci.edu}. For very faint hosts (typically for $m_g > 19.0$) with SDSS photometry and those hosts outside the SDSS footprint, we used our instrument SNIFS in imaging mode to obtain optical images in standard $ugriz$ filters. For some hosts, $g$-band photometry was obtained with Keck LRIS \citep{oke95} prior to spectroscopic observations of the hosts, and was later zero-pointed to either SDSS or SNIFS photometry. Magnitudes for our hosts in each photometric band were measured by performing common aperture photometry using SExtractor \citep{sextractor} in dual image mode, and corrected for foreground Milky Way reddening using the dust maps of \citet{sfd98} with the reddening law of \citet{ccm}.

% SPS method!
Stellar masses and SFRs were calculated from photometric data using a Bayesian method very similar to that presented in \citet[][and references therein]{salim07}. We generated a suite of model galaxies with physically-motivated star-formation histories, reddened by dust following a distribution derived empirically from the SDSS main galaxy sample. Central values and errors for \snia\ host galaxy mass-to-light ratios and specific SFRs (sSFR -- the SFR per unit mass) were calculated as the median and $\pm1\sigma$ values of the respective probability distribution functions for those quantities calculated from the weighted (by \chisq\ matching to photometry) values of the models. Total stellar masses were computed from the calculated mass-to-light ratios and the observed galaxy magnitude (here in $g$-band). Host sSFRs used in this analysis are the average over the past 0.5~Gyr. Our method was validated by examining both the ability to recover input galaxy parameters for our model galaxies, as well as with a sample of real galaxies from SDSS with galaxy mass and sSFR values measured by the MPA-JHU group \citep{kauff03a, brinchmann04}.  We found very good agreement in both tests, indicating our galaxy photometry fitting method is both accurate and consistent with previous studies.

%------------------
% spectroscopy
Gas-phase metallicities, specifically the logarithmic gas-phase oxygen abundance $12+\log(\mathrm{O/H})$, for hosts exhibiting star-formation were derived from emission line fluxes in galaxy spectra. Our science goals required the systemic host galaxy redshift as measured at the host galaxy core. Though the IFU capabilities of SNIFS sometimes resulted in galaxy signal being present in SNIFS data cubes, the small field of view (6\arcsec$\times$6\arcsec) required additional longslit spectroscopy observations to successfully include the galaxy core of most \snia\ hosts. Longslit spectra for our \snia\ host galaxies were obtained during numerous observing runs at multiple telescopes from 2007-2011. The instruments used were the Kast Double Spectrograph \citep{kast} on the Shane 3-m telescope at Lick Observatory, the Low Resolution Imaging Spectrometer \citep[LRIS --][]{oke95} on the Keck I 10-m telescope on Mauna Kea, the R-C Spectrograph 
%\footnote{http://www.ctio.noao.edu/spectrographs/4m\_R-C/4m\_R-C.html} 
on the Blanco 4-m telescope at Cerro Tololo Inter-American Observatory, the Goodman High Throughput Spectrograph \citep{clemens04} on the Southern Astrophysical Research (SOAR) 4-m telescope on Cerro Pachon, and GMOS-S \citep{davies97} on the Gemini-S 8-m telescope on Cerro Pachon. Additionally, some hosts had spectra available from SDSS DR8 \citep{aihara11}. In this work we use only longslit and SDSS host spectra, with all longslit spectra extracted from apertures dominated by light in the host core. While our host spectroscopy sample may have slightly different apertures, our focus on the galaxy core for all spectra enables a galaxy metallicity measurement consistent with large surveys such as SDSS. Details on instrument configurations, data reduction, and individual observations are presented in Paper~I.

% line fluxes and redshifts!
Emission line fluxes were measured by fitting each galaxy spectrum as a series of Gaussian emission line profiles superposed on a linear combination of simple stellar population (SSP) spectra from \citet{bc03} spaced  uniformly in logarithmic stellar age, following the same procedure as that of \citet{trem04}.
% - Metallicities
Emission line fluxes were corrected for internal reddening within the host galaxy by employing the Balmer decrement method \citep{agn2}, and galaxies whose emission line fluxes were contaminated by AGN activity \citep{bpt81, kewley06} were cut. Initial metallicities for ``high'' metallicity hosts (defined by $\log(\mathrm{NII}/\mathrm{H}\alpha) > -1.3$) were calculated using the ``N2'' method of \citet{pp04}, while for ``low'' metallicity hosts initial metallicities were calculated using the ``R23'' method of \citet{kk04}. All metallicities were then converted to the \citet{trem04} scale using the conversion formulae presented in \citet{ke08}. This enables us to use the strongest emission lines across the full range of metallicities while retaining a common scale (see discussion in Paper~I). The systematic uncertainty in such metallicity conversions is typically about 0.06~dex, so we have added this value in quadrature to the metallicity measurement errors for our \snia\ hosts.

%---------------------------------------------------------------
\subsection{\snia\ Light Curve vs. Host Properties}
\begin{figure*}
\begin{center}
\includegraphics[width=0.90\textwidth]{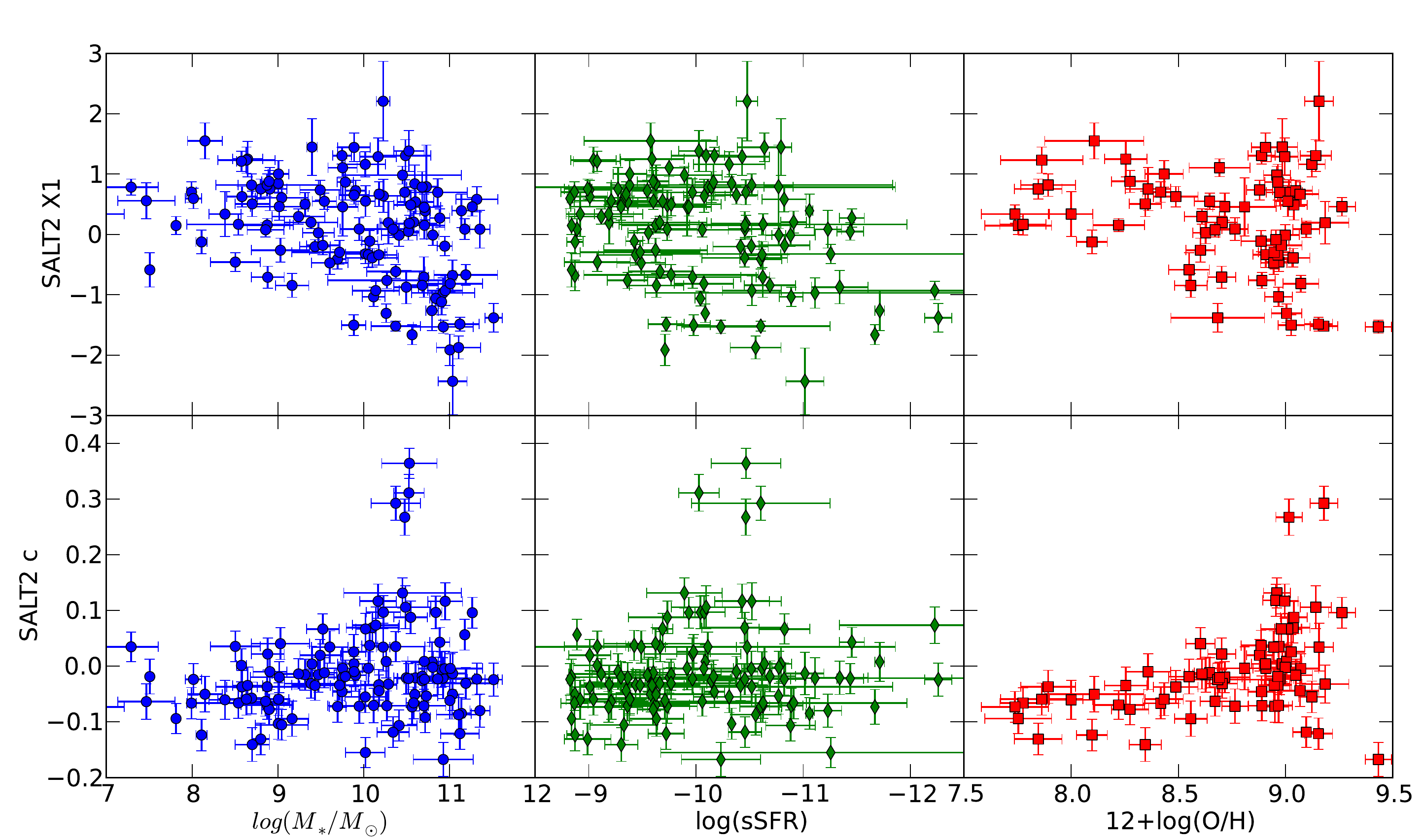}
\end{center}
\caption{Light curve width (SALT2 $x_1$ - top row) and color (SALT2 c - bottom row) for our \sneia\ as compared to the properties of their hosts: stellar mass (left column), specific star formation rate (middle column), and gas phase metallicity (right column).}
\label{fig:lc_vs_hosts}
\end{figure*}

% LC params vs. hosts!
We plot the \snia\ light curve parameters against the properties of their host galaxies in Figure~\ref{fig:lc_vs_hosts}. Trends of \snia\ properties with the properties of their host galaxies are important for detecting underlying physical mechanisms driving the shape of \snia\ light curves or their colors.

% STRETCH TRENDS
% history
Early studies of \snia\ host galaxies found qualitative evidence for a correlation between the observed peak magnitude (before correction for light curve width and color), light curve decline rate, and expansion velocity of an \snia\ with the morphological type of its host galaxy \citep{filippenko89, branchvdb93, hamuy96, gallagher05} such that brighter slower declining \sneia\ preferentially occur in later type (spiral and irregular) galaxies, while fainter faster declining \sneia\ preferentially occur in earlier type (elliptical and S0) galaxies. Recent analyses of large samples of \snia\ host galaxy masses have yielded the similar result that the observed brightnesses of \sneia\ correlate with the stellar masses of their host galaxy \citep[e.g.][]{howell09, neill09}, such that more massive hosts produce preferentially fainter \sneia.

% snf results: stretch vs. hosts
Our sample of \snf\ \sneia\ shows a similar trend of light curve widths with the stellar masses of their host galaxies. We also detect trends of light curve width with galaxy sSFR and metallicity. Such correlations are expected given that these quantities correlate with stellar mass in normal galaxies. What is perhaps most interesting about these figures is that rather than exhibiting a tight correlation of stretch with host mass or metallicity, the data instead indicate the presence of both high and low stretch \sneia\ in high mass (metallicity) hosts, while low mass (metallicity) galaxies seem to host exclusively intermediate to high stretch (i.e. brighter) \sneia.

% COLOR TRENDS
% history
% snf results: color vs. hosts
An examination of \snia\ colors in the \snf\ sample shows that very red \sneia\ (e.g. $c \gtrsim 0.2$) only occur in high mass (and high metallicity) hosts. This is perhaps to be expected as these \sneia\ are predicted to suffer from extinction by foreground dust, which is more abundant in high mass (and high metallicity) star-forming galaxies \citep{lee09, garn10}. Interestingly, we notice an apparent trend of \snia\ color with host metallicity, even excluding the apparently highly-reddened outliers. We will revisit this trend and its implications in Section~\ref{sec:host_bias_origin}.

%%%%%%%%%%%%%%%%%%%%%%%%%%%%%%%%%%%%%%%%%%%%%%%%%%%%%%%%%%%%%%%%%%%%%%%%%%%%%
\section{\snia\ Brightnesses and Host Galaxy Properties}
\label{sec:salt_trends}
Using the \snia\ and host galaxy data described above (see Section~\ref{sec:data}), we investigate the correlation of \snia\ host galaxy properties with the brightnesses of \sneia\ after the application of the standard stretch- and color-based luminosity corrections. In the first part of this Section we will use the SALT2 Hubble residuals, then examine the Hubble residuals derived using the spectroscopic luminosity correction method of \citet{bailey09} and their correlation with host properties.

Before beginning the analysis, we briefly discuss our procedure for weighting the data. We require that fitted models have reasonable goodness-of-fit, as judged using the value of $\chi^2$ per degree of freedom, $\chi^2_{\nu}$.  For most supernova Hubble residual analyses it has been necessary to add an extra intrinsic dispersion to obtain $\chi^2_{\nu}\sim1$ \citep[e.g.][]{conley11}. The SALT2 version~2.2 lightcurve model has an associated uncertainty that is propagated into the uncertainty on the Hubble residuals. For the \snf\ dataset the intrinsic uncertainty on the SALT2 color dominates the error budget, and leads to very similar uncertainties for all of our \sne. Ultimately, these uncertainties give $\chi^2_{\nu}\sim0.94$ (for 116 degrees of freedom), which is acceptably close to 1. The size of the observed \snia\ host mass bias is small compared to the observed scatter of \snia\ Hubble residuals and thus has little affect on the goodness-of-fit. In light of these considerations, for the analysis in this Section we do not add an intrinsic dispersion to our Hubble residuals when fitting linear models. However, for binned means we derive uncertainties based on the reduced weighted RMS (the weighted RMS divided by the square root of the number of degrees of freedom) for the data in each bin, which effectively accounts for the observed scatter of the data even if the uncertainties assigned to each SN were to be slightly misestimated.

%------------------
\subsection{Hubble Residuals vs. Galaxy Properties}
% general description, w/ sample sizes, etc.
In this analysis we use the Hubble residuals for 119 \sneia\ from \snf\ after corrections have been made for light curve width (SALT2 $x_1$) and color (SALT2 c). Of these 119 \sneia, 115 have host stellar mass and sSFR estimates from photometry, 2 are apparently hostless, and 2 are lacking sufficient host photometry data. Of the 119 \sneia\ considered here, a total of 69 (all of which are in the subset with host masses) have good gas-phase metallicities (i.e. are star forming and have emission line fluxes consistent with star formation rather than AGN activity -- see Section~\ref{sec:data} for details). 

\begin{figure}[ht]
\begin{center}
\includegraphics[width=0.45\textwidth]{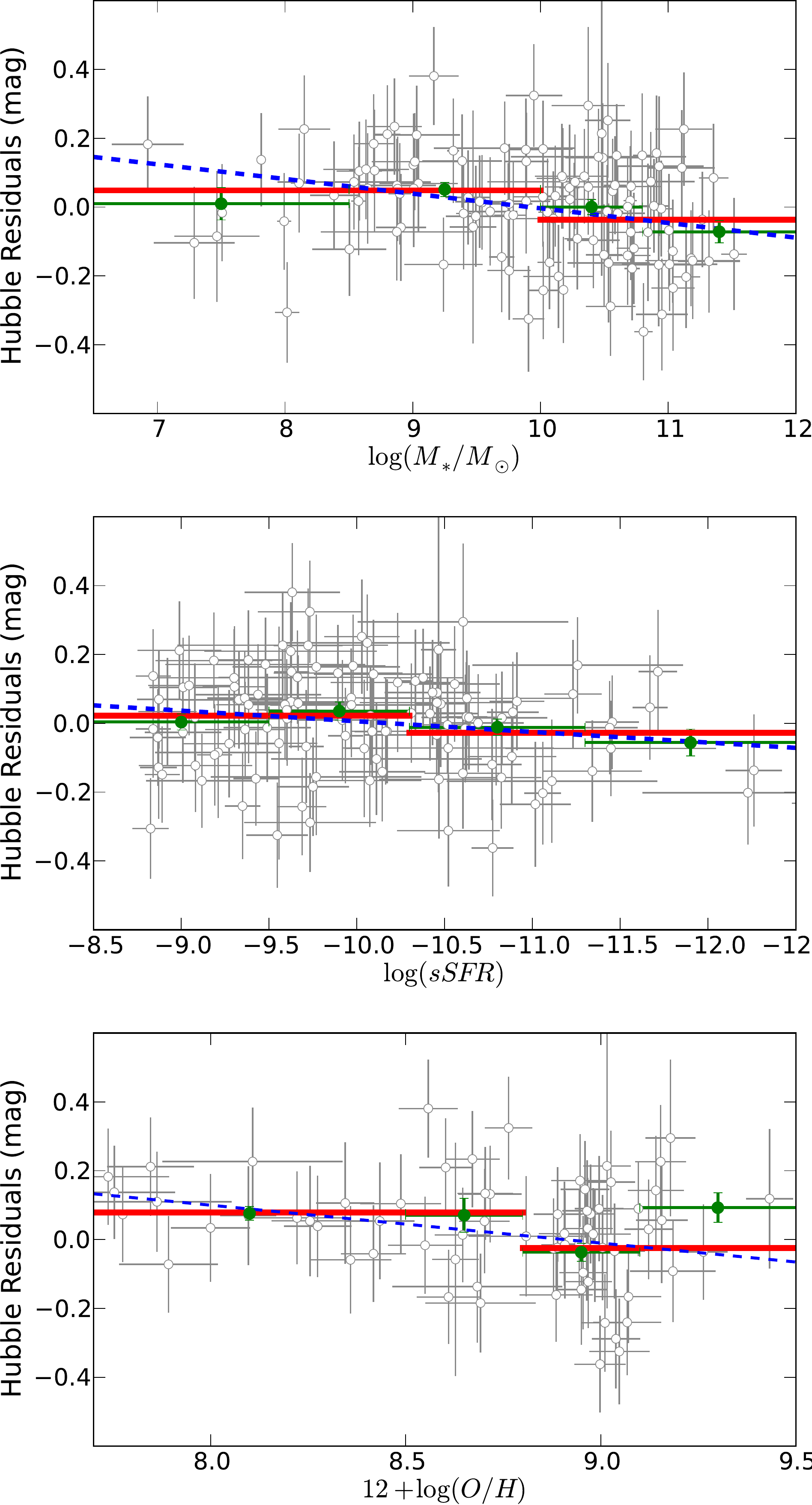}
\end{center}
\caption{Top: SALT2 Hubble residuals for \snf\ \sneia\ plotted versus host galaxy stellar mass (grey points). The blue line represents the best fit linear trend, the green points represent binned averages, and the thick red lines represent the averages for Hubble residuals split into high and low mass bins. Middle: Same as top, but for host sSFR. Bottom: Same as top, but for host gas-phase metallicity.}
\label{fig:dmu_x1c_vs_mass}
\end{figure}

In Figure~\ref{fig:dmu_x1c_vs_mass}
we plot the SALT2 Hubble residuals against host galaxy stellar mass, sSFR, and metallicity respectively. For visual aid we show the bin-averaged values of the \snia\ Hubble residuals in bins of host mass (sSFR, metallicity), as well as the best fit linear trend and mean residuals split by host mass (sSFR, metallicity).

% linear trends and splits!
Our data indicate a correlation of \snia\ Hubble residuals with all three host galaxy properties: mass, sSFR, and metallicity. We quantify this observed trend with two metrics: a linear trend in Hubble residuals vs. host stellar mass (sSFR, metallicity), and the difference between the average Hubble residuals when \sneia\ are split into two bins corresponding to high- and low-mass (sSFR, metallicity) hosts. The linear trend is fit by minimizing the sum of \chisq\ in the two-dimensional Hubble-residual host mass (sSFR, metallicity) plane via an orthogonal distance regression.  We note the slopes and \chisq\ values for this method are extremely close to a simple one-dimensional linear fit obtained by minimizing the Hubble residual \chisq, due to the very shallow slope of the trend. For this and following sections the uncertainty quoted on the ``step'' value corresponds to the quadrature sum of the uncertainties on the weighted Hubble residual mean values for each side of the step. We confirmed our confidence intervals on the step by randomly re-assigning host masses to Hubble residual values, and examining the distribution of step values from many realizations of this process.

% TABLE 1
\begin{table*}
\begin{center}
\caption{Hubble Residual Trends with Host Properties}
\label{tab:hubble_trends}
\begin{tabular}{cccrrrcrrc}
\hline
Host        & Residual       & $N_{SNe}$ & Linear Trend       & Linear  & Split & $N_{lo}$/$N_{hi}$ & Hubble Residual & Step    & HR Step      \\
Property    & Type           &          & (mag/dex)          & \chisq\ & Value &                  & Step (mag)      & \chisq\ & Significance \\
\hline
Mass        & Stretch+Color  & 115      & $-0.043 \pm 0.014$ &   100.5 & $ 10.0$ & $52$ / $63$ & $ 0.085 \pm 0.028$ &    98.8 & $3.0\sigma$ \\
sSFR        & Stretch+Color  & 115      & $ 0.031 \pm 0.008$ &   104.6 & $-10.3$ & $46$ / $69$ & $-0.050 \pm 0.029$ &   105.0 & $1.7\sigma$ \\
Metallicity & Stretch+Color  &  69      & $-0.106 \pm 0.043$ &    62.1 & $  8.8$ & $30$ / $39$ & $ 0.103 \pm 0.036$ &    60.0 & $2.9\sigma$ \\
\hline
$M_g$       & Stretch+Color  & 115      & $ 0.021 \pm 0.008$ &   101.5 & $-19.8$ & $57$ / $58$ & $-0.071 \pm 0.028$ &   101.6 & $2.5\sigma$ \\
$g-i$       & Stretch+Color  & 112      & $-0.059 \pm 0.030$ &   103.0 & $ 0.82$ & $56$ / $56$ & $ 0.069 \pm 0.029$ &   100.9 & $2.4\sigma$ \\
\hline
Mass        & Flux Ratio     &  94      & $-0.016 \pm 0.014$ &    92.4 &  $10.0$ & $38$ / $56$ & $ 0.050 \pm 0.028$ &    90.6 & $1.8\sigma$ \\
Mass        & Stretch+Color  &  94      & $-0.042 \pm 0.016$ &    77.7 &  $10.0$ & $38$ / $56$ & $ 0.066 \pm 0.030$ &    78.8 & $2.2\sigma$ \\
sSFR        & Flux Ratio     &  94      & $ 0.047 \pm 0.015$ &    85.4 &  $10.0$ & $39$ / $55$ & $-0.052 \pm 0.028$ &    87.6 & $1.9\sigma$ \\
sSFR        & Stretch+Color  &  94      & $ 0.021 \pm 0.017$ &    81.6 &  $10.0$ & $39$ / $55$ & $-0.038 \pm 0.030$ &    83.3 & $1.3\sigma$ \\
\hline
\end{tabular}
\end{center}
\end{table*}

We summarize the results of these fits in Table~\ref{tab:hubble_trends}. In this table we also present the \chisq\ values for the linear fits and split samples (which have two degrees of freedom less than the number of SNe for each subset). For reference, the \chisq\ of the Hubble residuals without any linear trend or step is 109.4 for the full sample of 115 \sneia, and 69.5 for the 69 \sneia\ with host metallicities. The linear fit of Hubble residuals against host mass has a significant slope. Likewise, the difference in Hubble residuals between high-and low-mass-hosted \sneia\ \citep[which we chose to split at $\log(M_*/M_\odot)=10.0$ following][]{sullivan10} shows a significant step in corrected \snia\ brightnesses. We find that low- and high-mass-hosted \sneia\ have brightnesses that differ by \saltmassmagstep\ \emph{after} stretch and color corrections have been applied, such that \sneia\ in high mass hosts are brighter after correction than those in low mass hosts. Our data also indicate that corrected \snia\ brightnesses differ for \sneia\ in hosts with different metallicity \citep[split at $12+\log(\mathrm{O/H})=8.8$ as in][]{dandrea11}. Linear or step-wise trends with star-formation intensity \citep[split at $\log(\mathrm{sSFR})=-10.3$ as in][]{sullivan10} are weaker, although consistent with previous studies. These results imply that applying the canonical \snia\ standardization techniques to \sneia\ at all redshifts will result in biased cosmological parameters as the \snia\ environments evolve in metallicity, and perhaps star-formation activity, as a function of redshift. We quantify this potential bias in Section~\ref{sec:host_corrections}.

%fits were typically of modest significance ($\sim 2\sigma$), 

\begin{figure}
\begin{center}
\includegraphics[width=0.45\textwidth]{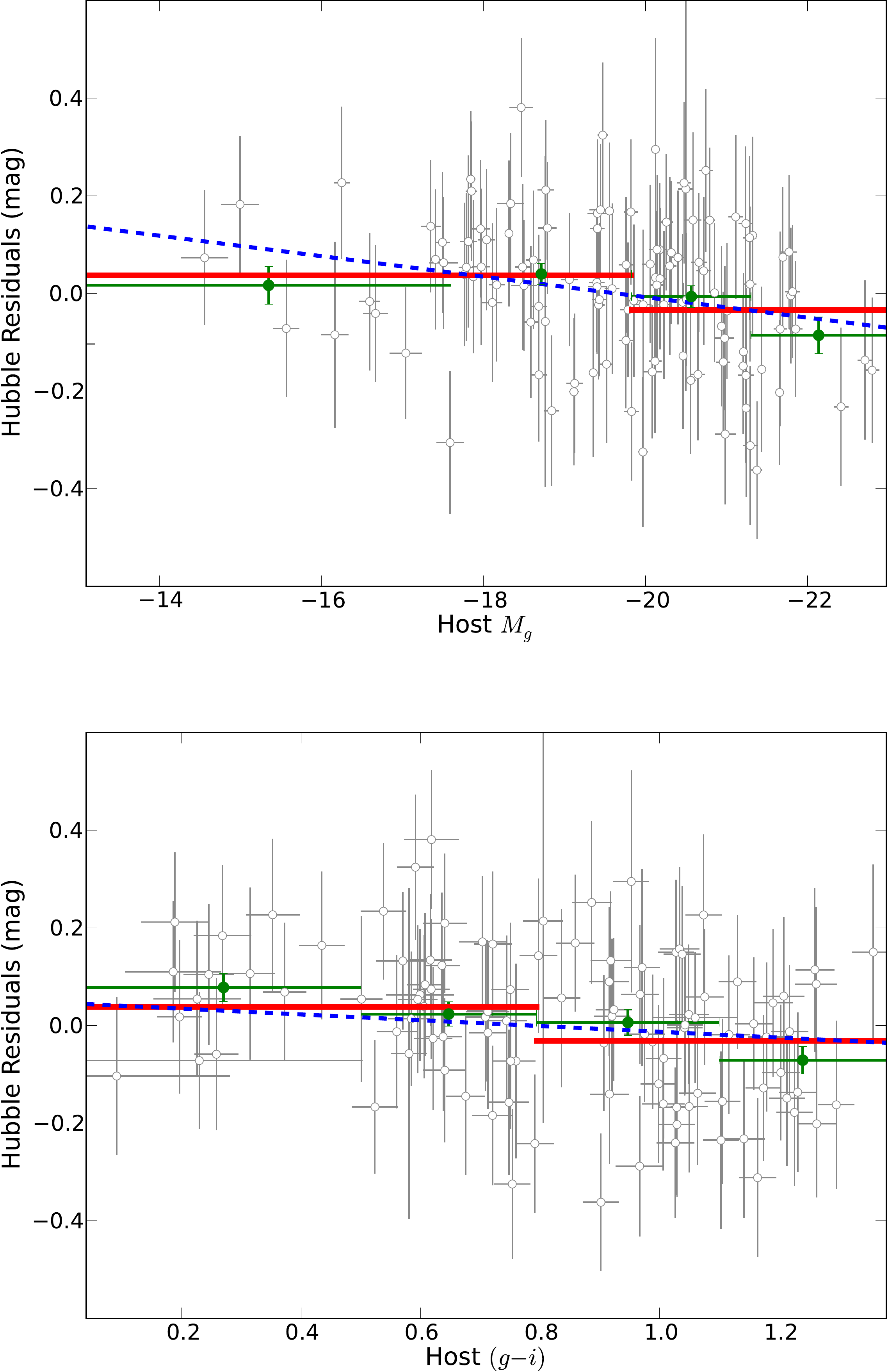}
\end{center}
\caption{Same as Figure~\ref{fig:dmu_x1c_vs_mass}, but with Hubble residuals plotted against model-free observational data: host absolute magnitude (in $g$-band, top panel), and host color ($g-i$, bottom panel).}
\label{fig:dmu_vs_gmag}
\end{figure}

% split by abs gmag and color!
To demonstrate the robustness of the observed host bias, we plot in Figure~\ref{fig:dmu_vs_gmag} the \snia\ Hubble residuals for our sample against two model-free observational quantities: the host galaxy absolute magnitude (in $g$-band), and the host galaxy color ($g-i$). Trends in \snia\ Hubble residuals with these properties are again evident, and we report the linear trend and magnitude steps (split at the median value of these quantities) for these host properties in Table~\ref{tab:hubble_trends}. Note that three \snia\ hosts did not have an $i$-band magnitude and thus do not appear in the sample with host $g-i$ colors. 

We note that these quantities are in the observer frame of reference and have not been K-corrected for the host galaxy redshift. Since this is a supernova-selected sample of galaxies, redshifts are uncorrelated with galaxy absolute magnitude or color and thus will not significantly alter the morphology of these plots. Furthermore, because of the low redshift range of our sample ($z \lesssim 0.1$), K-corrections were always less than 0.05~mag, comparable to the size of the errorbars in Figure~\ref{fig:dmu_vs_gmag}. Additionally, we have performed K-corrections for galaxy magnitudes (which are model-\emph{dependent}) and confirmed that the step in \snia\ Hubble residuals persists when using K-corrected photometry to derive host absolute magnitudes and colors. 

Thus we claim that the observed bias of \snia\ Hubble residuals with the properties of their host galaxies is not an artifact of erroneous modeling of host galaxy physical properties (indeed one would not expect such errors to correlate with supernova Hubble residuals in such a way as to produce a trend), but instead reflects a real physical relationship between the properties of \snia\ Hubble residuals with the nature of their environments.

We now discuss briefly why the observed trend of the corrected \snia\ Hubble residuals with host mass is very unlikely to be due to selection bias. One possible concern would be that in high mass galaxies \snia\ searches are only finding comparatively brighter \sneia\ due to the high background brightness from the massive host galaxy.  This is particularly concerning for high redshift surveys where the angular diameter of \snia\ host galaxies is comparable to the ground-based seeing element. When we consider the distributions of both host galaxy masses and \snia\ stretches (i.e. $x_1$) as a function of redshift for our \snf\ sample, we find no detectable change in the shape of these distributions as a function of redshift up to at least $z=0.12$. This is explored in more detail in Childress et al. (2012c, in prep.) where we examine the \snia\ host galaxy mass distribution. Here we note that this result implies that \sneia\ from \snf\ are not preferentially missed in high mass galaxies as the redshift increases and the host background thereby increases due to decreased angular size.

%------------------------------------------------------------------------
\subsection{Spectroscopically-Corrected \snia\ Brightnesses
and Host Properties}
\label{sec:bailey_trends}
We now revisit the above analyses with an alternate \snia\ luminosity standardization technique developed by \citet{bailey09}, which employs spectral flux ratios to standardize \sneia. The objective of this Section is to examine whether the host bias in corrected Hubble residuals persists when using a spectroscopic brightness correction technique, or whether it is only photometric standardization methods that suffer this bias.

The \snia\ data set for this analysis consists of the subset of the 119 \sneia\ from the previous sample for which we can apply the correction method of \citet{bailey09}. This method requires a spectrophotometric observation of the SN within $\pm2.5$ days of $B$-band maximum light (as estimated from the full light curve). This requirement brings the parent sample of \sneia\ down to 98, of which we have host stellar masses for 94. The measurement uncertainties using this technique are small, and the resulting scatter about the Hubble diagram revealed the presence of an intrinsic dispersion component.  Thus a blinded intrinsic dispersion was added in quadrature to the measurement errors in order to produce $\chi^2_\nu=1$ for the Hubble diagram.

\begin{figure}
\begin{center}
\includegraphics[width=0.45\textwidth]{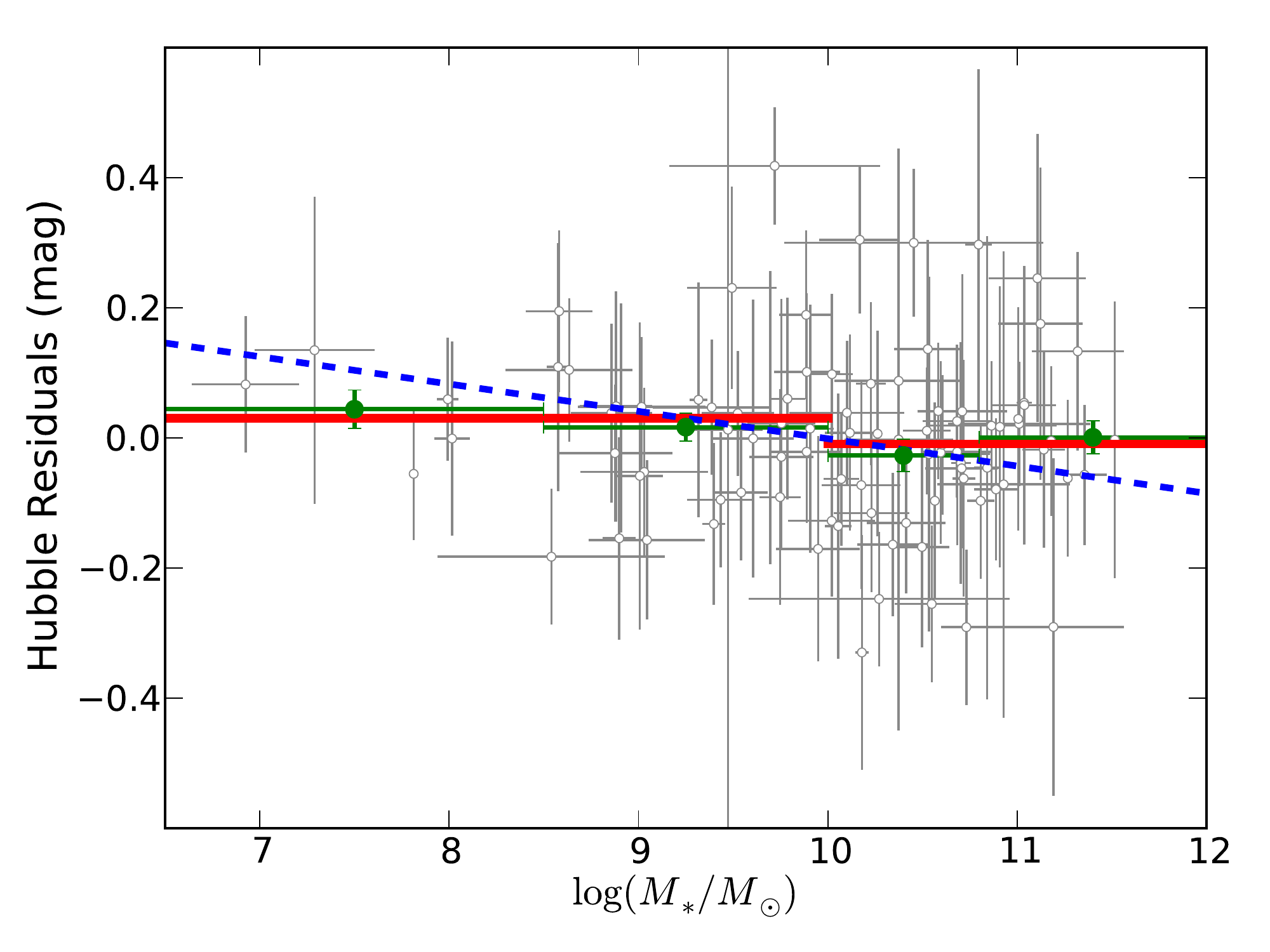}
\end{center}
\caption{Same as Figure~\ref{fig:dmu_x1c_vs_mass}, but with Hubble residuals obtained using spectral flux ratio brightness corrections following the method described by \citet{bailey09}.}
\label{fig:dmu_sjb_vs_mass}
\end{figure}

For simplicity in this section, we inspect only the trends of Hubble residuals with stellar mass and sSFR, since the sample attrition for a metallicity analysis is rather significant. In Figure~\ref{fig:dmu_sjb_vs_mass} we show the flux-ratio-corrected \snia\ Hubble residuals for our 94 \sneia\ plotted against the stellar masses of their host galaxies. Similarly to Figure~\ref{fig:dmu_x1c_vs_mass}, we plot the binned average Hubble residuals, best fit linear trend, and average Hubble residual when splitting the sample by host mass.

The results of our analysis show a trend with host mass similar to that found for stretch- and color-corrected Hubble residuals.
% step results
The high- to low-mass magnitude step after flux-ratio correction is \baileymassmagstep, versus \saltmassmagstep\ from stretch- and color-corrected \snia\ brightnesses. 
% linear trend results
The linear best-fit trend of flux ratio Hubble residuals with host mass is \baileymassmagslope, compared to \saltmassmagslope\ for stretch- and color-based Hubble residuals for the same sample.

% comparison on common sample
While at first glance this appears to be a decrease in the observed host bias, the improvement is less marked when comparing the bias on common \snia\ subsamples. Since those \sneia\ without a spectroscopic observation near maximum light were excluded from this sample, we repeat the original calculation of the previous section for stretch- and color-corrected Hubble residuals for the same sample of \sneia\ used in the flux-ratio analysis. Doing so we find the luminosity step for this sample is \saltsjbmassmagstep, and the slope of the linear trend fit is \saltsjbmassmagslope. These results are summarized in Table~\ref{tab:hubble_trends} along with the rest of our Hubble residual trend results.

The result is similar when considering trends in sSFR. The flux-ratio Hubble residual step in sSFR for this sample is \baileyssfrmagstep\ versus \saltsjbssfrmagstep\ for the same sample with stretch- and color-corrected Hubble residuals. Similarly, the slope of flux-ratio Hubble residuals with sSFR is \baileyssfrmagslope\ versus \saltsjbssfrmagslope\ for the same sample of \sneia\ with SALT2 Hubble residuals. Thus the flux-ratio-corrected Hubble residuals show a host bias which is consistent with that found using standard stretch and color corrections to \snia\ luminosities.

%%%%%%%%%%%%%%%%%%%%%%%%%%%%%%%%%%%%%%%%%%%%%%%%%%%%%%%%%%%%%%%%%%%%%%%%%%%%
% COMBINED DATA SET RESULTS
\section{\snia\ Hubble Residual Host Bias Results from Multiple Data Sets}
\label{sec:combined_data_sets}
Our work reinforces the results of recent \snia\ host studies which all have found that the stretch- and color-corrected luminosities of \sneia\ are biased with respect to the properties of their host galaxies. Such a trend is quite distressing for future \snia\ cosmology missions which will hunt for \sneia\ at very high redshifts where the mean stellar age and metallicity are quite different from their values in the local universe. The discovery of this \snia\ host bias has motivated two critical questions: what is the true cause of this bias, and what should be done to correct it? We will address each of these questions in turn in Sections~\ref{sec:host_bias_origin} and \ref{sec:host_corrections}, but first we analyze the impact of our \snf\ data on their answers. In this Section we begin by comparing the results of our analysis with those of previous studies, then we inspect the strength of the observed \snia\ host bias when combining data from multiple surveys. 

%---------------------------------------------
\subsection{Comparison To Previous Studies}
In this section we summarize the results presented here and compare those to the findings of previous studies. We compare our results to previous studies with respect to host stellar mass \citep{kelly10, sullivan10, lampeitl10, gupta11}, then compare our trend with host metallicity to those of \citet{konishi11} and \citet{dandrea11}. In Table~\ref{tab:host_bias_results} we summarize the results of these previous studies in terms of the Hubble residual slope and step values, as well as their consistency with the results derived in this work for \snf. For reference we also show the results for the combined sample analysis outlined below in Section~\ref{sec:combined_HR_results}.  

% TABLE 2 - host studies summary
\begin{table*}
\begin{center}
\caption{Survey of SN~Ia Host Bias Results}
\label{tab:host_bias_results}
\begin{tabular}{lccccccccc}
\hline
Authors & SN     & Host     & $N_{SNe}$ & Median   & Median     & Linear Trend               & \snf\     & Hubble Residual             & \snf\     \\  
        & Survey & Property &          & Redshift & $M$ or $Z$ & (mag/dex)$^a$ & Agreement & Step (mag)\tablenotemark{a} & Agreement \\
\hline
Sullivan '10 & SNLS  & M & 195 & 0.64 &  9.8 & $-0.042\pm0.013$ & $0.05\sigma$ & $0.08 \pm0.02 $     & $0.15\sigma$ \\
Lampeitl '10 & SDSS  & M & 162 & 0.22 & 10.3 & $-0.072\pm0.018$ & $1.27\sigma$ & $0.100\pm0.025$     & $0.40\sigma$ \\
Gupta '11    & SDSS  & M & 206 & 0.22 & 10.3 & $-0.057\pm0.019$ & $0.59\sigma$ & $0.096\pm0.028$     & $0.28\sigma$ \\
Kelly '10    & Low-z & M &  62 & 0.03 & 10.7 & $-0.15 \pm0.06 $ & $1.74\sigma$ & $0.094\pm0.045^b$ & ---          \\
This Work    & SNf   & M & 115 & 0.03 & 10.1 & $-0.043\pm0.014$ & ---          & $0.085\pm0.028$     & ---          \\
This Work    & All   & M & 601 & 0.25 & 10.2 & $-0.037\pm0.011$ & $0.34\sigma$ & $0.077\pm0.014$     & $0.26\sigma$ \\
\hline
Konishi '11  & SDSS  & Z &  72 & 0.22 &  8.9 & ---              & ---          & $0.13 \pm0.06 $     & $0.39\sigma$ \\
D'Andrea '11 & SDSS  & Z &  34 & 0.22 &  8.8 & ---              & ---          & $0.114\pm0.053^c$ & $0.16\sigma$ \\
This Work    & SNf   & Z &  69 & 0.06 &  8.9 & $-0.106\pm0.043$ & ---          & $0.103\pm0.036$     & ---          \\
\hline
\end{tabular}
\end{center}
$^a$Values as originally published by stated authors.\\
$^b$Split at $\log(M_*/M_\odot)=10.8$\\
$^c$Including added intrinsic scatter, see text.
\end{table*}

Two items in this table are worth noting here. The first is that the Hubble residual step (with respect to host mass) quoted in \citet{kelly10} comes from splitting that sample at $\log(M_*/M_\odot)=10.8$, rather than at 10.0 as was done for all other sample, and so is not directly comparable to our \snf\ result. The second item of note regards the uncertainty in Hubble residual step (with respect to metallicity) for the \citet{dandrea11} results. These authors reported a step value of $\Delta m_{corr} = 0.114 \pm 0.023$~mag, but the error bar on this result is derived from Hubble residuals whose uncertainties were not adjusted to give an acceptable goodness of fit. This choice was an attempt to avoid deweighting the data in the event most of the remaining scatter arose from host-dependent bias. However, the linear and step models remove little of the excess scatter in the data presented by \citet{dandrea11} so their quoted uncertainty underestimates the true scatter in the data. Given their sample size, we estimate that an intrinsic scatter term of $\sigma_\mathrm{int}=0.14$~mag \citep[cited by][as the scatter in their Hubble residuals]{dandrea11} would make the uncertainty in their result increase to 0.053~mag.

We see from Table~\ref{tab:host_bias_results} that the \snf\ results are fully consistent with previous studies of the \snia\ host bias. The only major exception is the disagreement of the \citet{kelly10} Hubble residuals slope with both our results and those of other authors. The \citet{kelly10} sample covered a shorter mass range than the other three samples, and we will show in the following section that their result is consistent with detailed structure that we have uncovered in the trend of \snia\ Hubble residuals over their shorter host mass range.

%---------------------------------------------
\subsection{Hubble Residual Host Trends from Combined Data Sets}
\label{sec:combined_HR_results}
Our analysis of \snf\ \snia\ Hubble residual trends with host galaxy properties adds to the continually growing body of data that indicates current \snia\ standardization methods produce a bias of Hubble residuals with respect to host galaxy properties. We now examine the observed bias when multiple data sets are combined, in part to address any concern that such a bias could be the result of spurious effects or inaccurate galaxy stellar population synthesis methods. Here we will show that the combination of multiple data sets reveals detailed structure of the Hubble residual trend with host mass.

For this purpose, \snia\ Hubble residuals and host galaxy data have been collected from two additional untargeted \snia\ searches: the Supernova Legacy Survey \citep[SNLS -- ][]{sullivan10, guy10} and the SDSS-SN survey \citep{gupta11}, as well as the compilation of low-redshift literature \sneia\ and their hosts of \citet{kelly10}. We apply a common flat $\Lambda$CDM cosmology with $\Omega_M=0.271$ \citep{sullivan11a, suzuki12} and \snia\ standardization parameters $\alpha=0.139$ and $\beta=3.1$ \citep{guy10}.
We note that these \snia\ luminosity correction parameters are appropriate for the data here because they are derived for the same version (2.2) of the SALT light curve fitter used to fit all the data, and were derived without fitting simultaneously for a host mass correction term.

% REWRITE 2012-07-02
% data set sources
% 1 - SNLS
Host galaxy masses for the SNLS sample were derived by \citet{sullivan10}, and \snia\ light curve parameters 
% ($x_1$, $c$, and $m_B$) 
and redshifts were presented in \citet{guy10}. Hubble residuals for this sample were derived directly using the aforementioned cosmology and \snia\ standardization parameters.
% 2 - SDSS
Host galaxy data and \snia\ Hubble residuals for the SDSS-SN sample were presented in \citet{gupta11}. We applied small corrections to their reported Hubble residuals and errors in order to convert from their cosmology and light curve parameters \citep{kessler09, holtzman08} to ours.
% 3 - Kelly
Host masses for the literature low-redshift sample were derived in \citet{kelly10}, while light curves were presented in \citet{hicken09a}. We use new Union3 (Rubin et al. 2013, in prep.) light curve fits for 61 of the 62 \sneia\ from the \citet{kelly10} sample using version 2.2 of SALT, omitting the slow decliner SN~2006bz. These updated light curve fits were performed in order achieve consistency with the three other samples, as the updated SALT color model behaves differently than that used in the light curve fits from \citet{kelly10}.
% 4 - SNf
For \snf\ data, we use the raw blinded Hubble residuals (before stretch and color correction) then apply corrections for the chosen \snia\ standardization parameters.

% Host masses are compatible!
We note that the \citet{sullivan10} and \citet{gupta11} host masses are both computed using a \citet{kroupa01} IMF, whereas our values are computed with a \citet{chab03} IMF. Fortunately these IMFs are extremely similar, and while they may exhibit some relative scatter (perhaps 0.05~dex) they do not require an offset in mass-to-light ratio zeropoints. The host galaxy sample of \citet{kelly10} uses masses computed with the \citet{ranabasu92} IMF, which is the default IMF in \zpeg\ \citep{zpeg}. From a sample of approximately 10,000 galaxies from SDSS, we computed \zpeg\ masses using both the \citet{ranabasu92} and \citet{kroupa01} IMFs and found a mean offset of 0.024~dex (which we note is much smaller than typical mass uncertainties of 0.1~dex), which we subtract from the \citet{kelly10} masses for consistency with our IMF and that of SNLS and SDSS. Our host masses and those reported by the aforementioned authors were all computed using the same flat $\Lambda$CDM cosmology with $\Omega_M=0.3$ and $h=0.7$, indicating our mass scales are all consistent.

\begin{figure*}
\begin{center}
\includegraphics[width=0.85\textwidth]{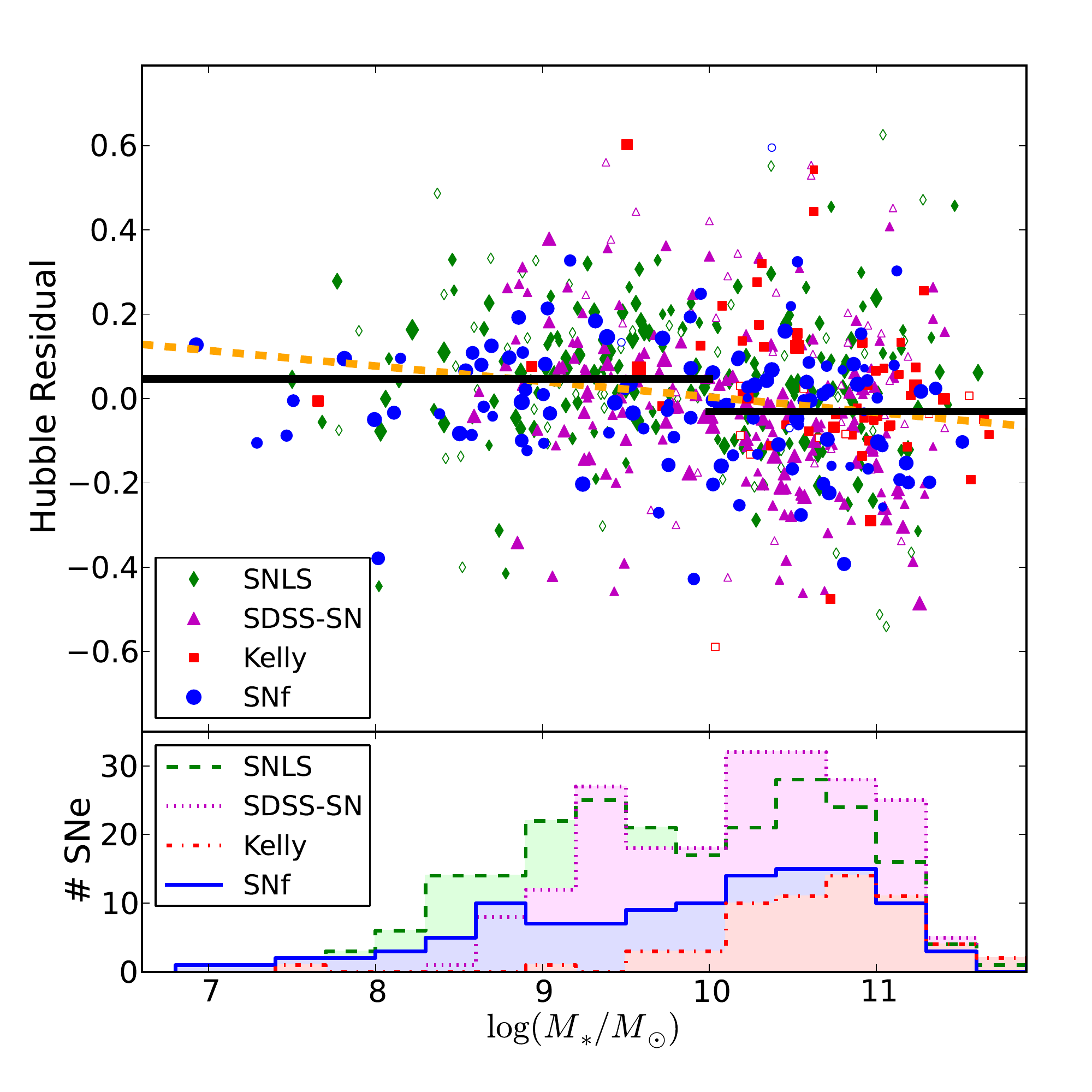}
\end{center}
\caption{Top: Combined Hubble residuals and host galaxy masses for the \snf\ sample (this work), the SNLS survey \citep{sullivan10, guy10}, the SDSS-SN survey \citep{gupta11}, and literature low-$z$ sample \citep{kelly10,hicken09a}; marked as blue circles, green diamonds, magenta triangles, and red squares, respectively. Symbol size is inversely proportional to Hubble residual error, and open symbols represent \sneia\ whose Hubble residual {\em measurement} errors exceed 0.2 magnitudes. The weighted mean Hubble residuals for high- and low-mass-hosted \sneia\ are shown as thick black bars, and the best fit linear trend is the thick dashed orange line. Bottom: Histograms of host mass distributions for the four samples.}
\label{fig:combined_dmu_vs_mass}
\end{figure*}

The combined sample of Hubble residuals as a function of host galaxy mass is presented in Figure~\ref{fig:combined_dmu_vs_mass}. The total number of \sneia\ included in this analysis now totals 601 (219 SNLS, 206 SDSS-SN, 61 literature low-$z$, and 115 \snf). 
% CHI2 TEXT
We now wish to calculate the step value and best fit linear trend from this combined data set, but once again must consider the proper weighting of the data.
% adapted from old text
The measurement uncertainties alone yield $\chi^2_\nu \sim 2$ for the combined sample. Thus, to obtain $\chi^2_\nu\sim1$ 
%as required for an acceptable model 
we add an intrinsic dispersion term of 0.10~mag in quadrature with the measurement uncertainties for all samples in the combined dataset, which yields a favorable \chisq\ of 628 for the sample of 601 \sneia. 
%As was found for the \snf\ data set in Section~\ref{sec:salt_trend}, inclusion of this intrinsic dispersion term had a minimal effect on the magnitude and significance of the host bias step because we use a reduced weighted RMS to assess the uncertainty of the steps.

% combined HR slope and step
With this combined data set we fit for a linear trend using the same two-dimensional \chisq\ minimization method employed for the \snf\ linear fits in Section~\ref{sec:salt_trends}. This is shown as the thick dashed orange curve in Figure~\ref{fig:combined_dmu_vs_mass}, and has a slope of $-0.037\pm0.011$~mag/dex. We then calculate the step in average Hubble residuals between \sneia\ in high- and low-mass hosts when splitting the sample in host mass at $\log(M_*/M_\odot)=10.0$ (this results in $N_{low}=255$ and $N_{high}=346$ \sneia\ in the two bins). We find a combined offset in corrected \snia\ brightnesses of \allmassmagstep, a $5.6\sigma$ detection of host bias in \snia\ Hubble residuals.

% HR step for each subset
% Sullivan 10 reproduced using empirical SALT2.2->SALT2.1 conversions
Concurrent with the measurement of the Hubble residual step for the combined \snia\ sample, we also calculate the step for SNLS and SDSS separately to compare against previously published results. We find the Hubble residual step is $0.065\pm0.020$~mag for SNLS and $0.114\pm0.025$~mag for SDSS-SN. We note that the step value for SNLS we derive here is smaller than that reported in \citet{sullivan10}, and we confirmed that this is due to the use of version 2.1 of the SALT light curve fitter in \citet{sullivan10}. Finally we checked the Hubble residual step for the \citet{kelly10} sample split at their value of $\log(M_*/M_\odot)=10.8$ using our chosen cosmology and found this step to be $0.108\pm0.045$~mag, consistent with their reported results.

%---------------------------------------------
\subsection{Structure in the Host Mass Hubble Residual Trend}
\label{sec:hr_trend_structure}
\begin{figure}
\begin{center}
\includegraphics[width=0.45\textwidth]{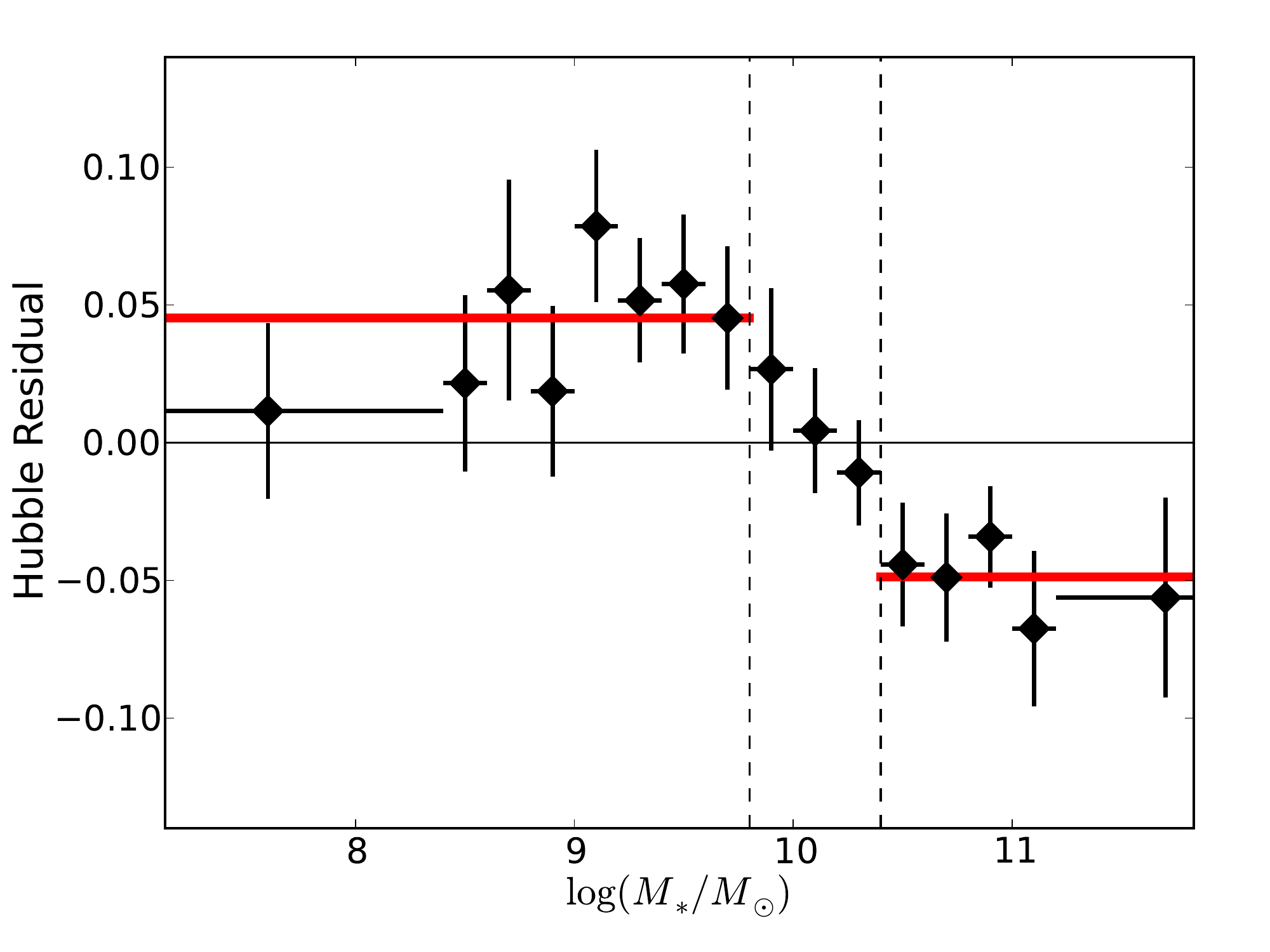}
\end{center}
\caption{Binned average Hubble residuals from Figure~\ref{fig:combined_dmu_vs_mass} in host mass bins of width 0.2~dex. The highest and lowest mass bins are wider in order to have at least 20 SNe in each bin. Error bars for each bin are the weighted RMS divided by the square root of $N-1$ where $N$ is the number of points in that bin. An apparent region of Hubble residual transition from $\log(M_*/M_\odot)=9.8$ to $\log(M_*/M_\odot)=10.4$ is denoted by the vertical dashed lines, while the mean Hubble residuals of \sneia\ in hosts of masses above and below this region are denoted by the thick red bars.}
\label{fig:dmu_mass_fine_bins}
\end{figure}

% 3 - STRUCTURE IN THE TREND!!!
We now use the combined \snia\ host mass and Hubble residual sample to inspect the structure of the trend of \snia\ Hubble residuals with host properties. We bin the sample in narrow bins of 0.2~dex in host galaxy stellar mass, and plot the result in Figure~\ref{fig:dmu_mass_fine_bins}. The result of this analysis indicates that  rather than a clear linear trend or step in host mass, the binned \snia\ Hubble residuals appear consistent with a plateau at both low and high stellar mass separated by a narrow transition region from $\log(M_*/M_\odot)=9.8$ to $\log(M_*/M_\odot)=10.4$.

To further illustrate this result, we recalculate the weighted mean Hubble residual for all \sneia\ with hosts more massive than $\log(M_*/M_\odot)=10.4$, and find that the difference between this value and the weighted mean Hubble residual for \sneia\ in hosts less massive than $\log(M_*/M_\odot)=9.8$ is \allwidemassmagstep. This value is larger than the canonical value we calculated from our split of all \sneia\ at host mass of $\log(M_*/M_\odot)=10.0$. This indicates that the Hubble residual host bias observed in the data is not best represented by a sharp step function and that previous measurements of the offset between \snia\ luminosities in high and low-mass hosts may be underestimating the true value of the offset.

We note that this result is not as clearly evident when combining the corrected Hubble residuals from the four samples without first applying a common cosmology and \snia\ correction parameters ($\alpha$ and $\beta$), nor is it as pronounced if any one of the \snia\ samples from untargeted searches (SNLS, SDSS-SN, SNf) is excluded. This result then is found only with the most up-to-date and comprehensive sample of \snia\ host galaxy masses and Hubble residuals.

This result also elucidates the likely cause of the strong trend of Hubble residuals with host mass reported by \citet{kelly10}, which we previously noted as being much stronger than that reported by other studies including our own. If most of the Hubble residual bias between low- and high-mass hosted \sneia\ is driven by a rapid shift in mean Hubble residuals over a short range, then the local Hubble residual gradient can be quite strong. Indeed if the entire offset of \allwidemassmagstep\ transitions over a mass range of 0.6~dex, then one would expect a gradient of \allwidemassgrad, which is remarkably close to the trend reported by \citet{kelly10}. Most of their \snia\ hosts were in the mass range of this transition, and thus serendipitously occurred in the region where the host bias is strongest.

For completeness here, we note that the weighted mean and RMS values have only a mild dependence on how one chooses to weight individual \sneia. For this Section and the analysis of \snf\ data (Section~\ref{sec:salt_trends}) weights were assigned according to the measurement errors alone, but our use of the reduced weighted RMS in assessing the uncertainty absorbed any unaccounted-for scatter. An alternative approach is to add an intrinsic dispersion in quadrature with the measurement uncertainties. This preserves the relative order of weighting of \sneia\ but decreases the difference in weights between \sneia\ with high and low measurement error. We have confirmed that adding an intrinsic dispersion term of 0.10~mag to the measurement uncertainties for all \sneia\ results in negligible ($\lesssim0.5\sigma$) change to the weighted mean and RMS values reported here.

%%%%%%%%%%%%%%%%%%%%%%%%%%%%%%%%%%%%%%%%%%%%%%%%%%%%%%%%%%%%%%%%%%%%%%%%%%%%%
% ORIGIN OF TREND
\section{Possible Origins of \snia\ Hubble residual Host Bias}
\label{sec:host_bias_origin}
The first major question raised by the \snia\ luminosity host bias is, what is its underlying physical cause? Is it a true correlation of \snia\ brightnesses with the properties of their progenitors? If so, is the trend being driven by progenitor age or metallicity (or possibly both), or perhaps some other physical property of the progenitor? On the other hand, could this be a deficiency in the \snia\ standardization techniques? In this Section we address these questions by comparing the observed \snia\ host bias to a series of physical models. Specifically we seek a physical model which can reproduce the fast transition of Hubble residuals over a short mass range discovered when we combined multiple data sets (see Section~\ref{sec:combined_data_sets}).

% describe my ideas...
One candidate driver of the host bias in \snia\ luminosities is erroneous color correction resulting from the interplay of \snia\ intrinsic color variation and reddening by foreground dust (Section~\ref{sec:two_color_effects} and Figure~\ref{fig:color_mc_example}). Our \snf\ data indicated a possible trend of \snia\ intrinsic color with host metallicity (Section~\ref{sec:color_metal_trend} and Figure~\ref{fig:salt2_color_vs_metal}), which we show could drive a trend of Hubble residuals with host properties. We then discuss the likely shift of average \snia\ progenitor ages across the range of host galaxy masses (Section~\ref{sec:age_based_host_bias} and Figure~\ref{fig:mass_dist_fit}) and its potential for driving some of the observed host bias. Finally we consider the effect of neutronization in \sneia\ \citep{tbt03, bravo10} and its effect on metallicity-dependent stretch corrections (Section~\ref{sec:bravo_models} and Figure~\ref{fig:bravo_models}).

Our findings are then summarized in Section~\ref{sec:models_summary}, where we plot in Figure~\ref{fig:dmu_trend_with_models} the observed trend of \snia\ Hubble residuals with host galaxy mass (see Section~\ref{sec:combined_data_sets}) against the trends predicted from all the models presented in this Section.

%------------------------------------
% 2-color effect
\subsection{Competing Effects of Intrinsic \snia\ Color and Dust}
\label{sec:two_color_effects}
A potential culprit that could drive the observed trend of \snia\ Hubble residuals with host properties is the correction of \snia\ luminosities for color. If there is significant variation in the \emph{intrinsic} colors of \sneia, especially if intrinsic color affects the \snia\ brightness differently than dust (i.e. different $\beta$), then correction of \snia\ brightnesses via a single color law will produce erroneous values. The magnitude and sign of the color correction error for each \snia\ is then related to the relative proportion of dust and intrinsic color variation. This can produce trends with host galaxy properties due to the correlation of galaxy dust content with mass \citep{lee09, garn10} such that more massive star-forming galaxies are more dusty. The magnitude and direction of the \snia\ color correction error is dependent on the slopes of the SN and dust color laws.

% DETAILED EXAMPLE
% A - intrinsic color
Here we illustrate this effect with a specific example utilizing Monte-Carlo simulations of mock \sneia. A key input to to simulation is the \snia\ intrinsic color distribution, for which we choose a Gaussian of width 0.07~mag \citep[e.g.][]{jha07, folatelli10}. We further assume that the intrinsic color affects the \snia\ luminosity following some slope $\beta_\mathrm{SN}=5.0$ such that a \snia\ which is intrinsically $\Delta c=0.02$~mag redder would be $\Delta m = \beta_\mathrm{SN}\Delta c = 0.10$~mag fainter.

% B - host dust
We generate a mock sample of \sneia\ whose intrinsic colors follow this distribution and are found in galaxies whose masses are distributed similarly to the combined sample of \sneia\ from Section~\ref{sec:combined_data_sets}. Utilizing the observed correlation of \snia\ stretch with host mass (see e.g. Figure~\ref{fig:lc_vs_hosts}), we assign a stretch to each mock SN. 

Finally we assign a host galaxy reddening to each \snia\ using the correlation of dust content with host mass in star-forming galaxies. The highest mass mock SN hosts are considered to be passive dust-free ellipticals with a likelihood of being passive trained on SDSS data. Then a foreground dust reddening of the SN was drawn from an exponential distribution scaled to the total host galaxy dust content. This critial piece accounts for the fact that an SN is likely to be geometrically distributed somewhere within the dust of its host galaxy. For this example we choose a dust reddening law of $\beta_\mathrm{dust}=2.8$ (i.e. $R_V=1.8$).

\begin{figure}
\begin{center}
\includegraphics[width=0.45\textwidth]{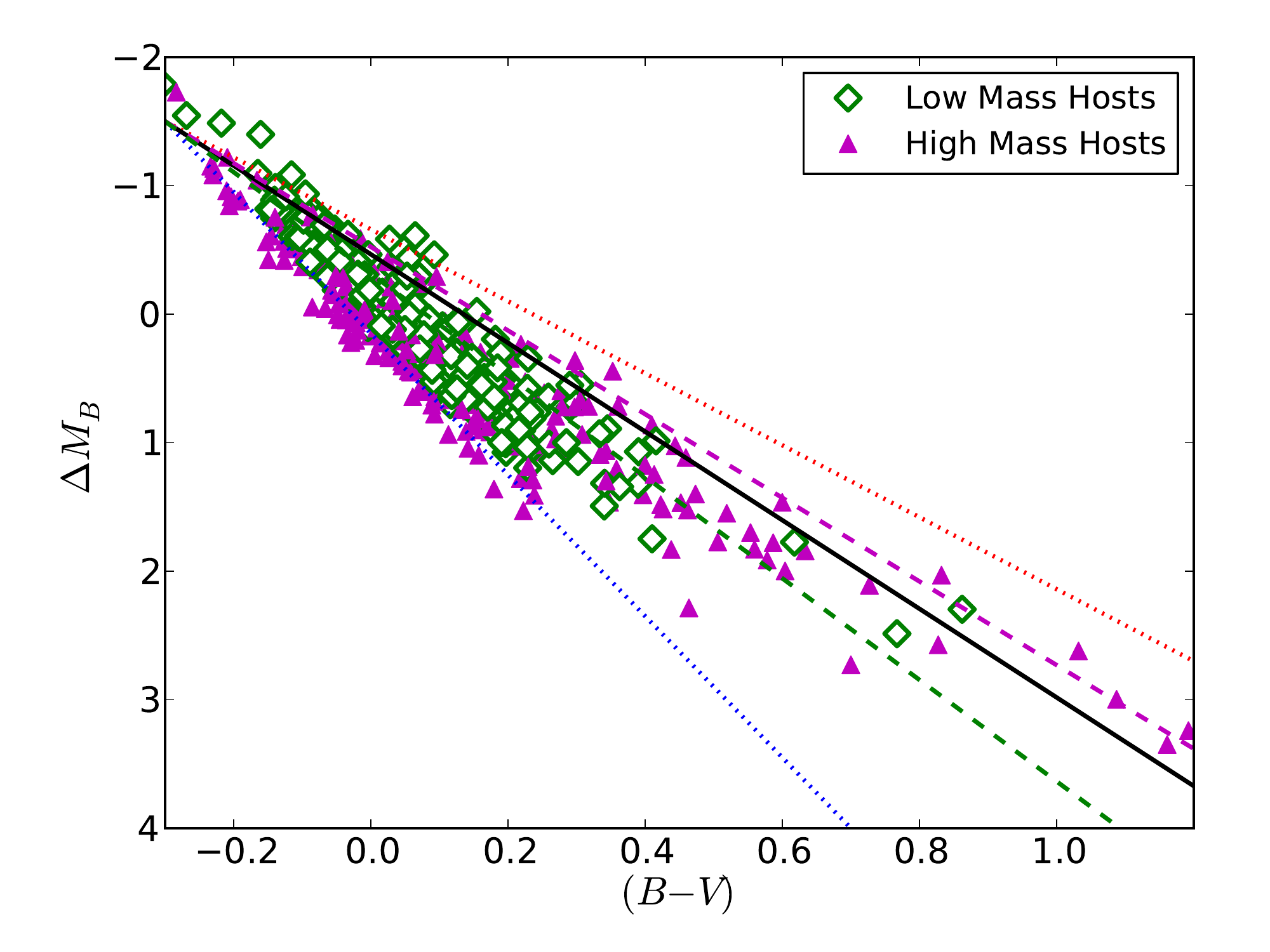}
\end{center}
\caption{Example Monte-Carlo simulation of competing \snia\ color variation of the \snia\ luminosity (steep dotted blue line) and foreground reddening by dust (shallow dotted red line). A fitted monolithic color law (black line) fits a slope between the two input slopes, while a color law fit split by host mass fits a color law (shallow dashed magenta line) for \sneia\ in high-mass hosts (filled magenta triangles) closer to the dust color law and a color law (steep dashed green line) for \sneia\ in low-mass hosts (open green diamonds) that is closer to the intrinsic \snia\ color law. All color law lines have been fixed to a common intercept to better illustrate the differing slopes.}
\label{fig:color_mc_example}
\end{figure}

From these components we then simulate the final color and luminosities of the mock \sneia, and these are shown in Figure~\ref{fig:color_mc_example}. Now we can see that a single monolithic color law fitted to this data will fit a line (solid black line) with slope intermediate between the input dust (shallow red dotted line) and SN intrinsic (steep blue dotted line) color laws. The luminosity correction error then depends on which side of the line the SN falls. Because dust content is greater (on average) in more massive \snia\ host galaxies, more of these \sneia\ (magenta triangles) will fall above the line and will be over-corrected for dust (and vice versa for \sneia\ in low mass galaxies, green diamonds). Similarly, if we split the sample in host mass and fit separate color laws, the high mass hosted \sneia\ have a lower $\beta$, closer to the input dust law, while low mass hosted \sneia\ have a higher $\beta$, closer to the input \snia\ intrinsic color law. This result in consistent qualitatively with that found by \citet{sullivan10} when splitting the SNLS SN sample by host mass.

% [FIGURE: example color law fit]

%(later):
This set of input parameters ($\beta_\mathrm{dust}$ and $\beta_\mathrm{SN}$) was purposely chosen for this example in order to produce the desired sign of the \snia\ host bias. A more detailed exploration of this two-color effect across the whole space of input parameters will be presented in Paper~III. For this work we focus on the trend of \snia\ Hubble residuals with host mass predicted to be produced by this effect. We thus generate a large sample (to reduce Monte Carlo noise) of mock \sneia\ using the above parameters, and the calculated Hubble residual bias curve is shown in Figure~\ref{fig:dmu_trend_with_models} of Section~\ref{sec:models_summary} (dotted magenta curve).

Despite the favorable qualitative results from this simulation, this two-color effect failed to produce sufficiently large offsets between low- and high-mass-hosted \snia\ Hubble residuals to match observations. Even the steep \snia\ color law used here produced a step size of about 0.03~mag. Similarly, the structure does not match that observed in the data (Section~\ref{sec:hr_trend_structure}), particularly at high host masses where the dust-free elliptical hosts dominate and the dust over-correction effect disappears. However, while this effect alone cannot explain the \snia\ host bias, the correlation of dust content with host mass is certain to be impacting the observed colors of \sneia\ and may still be at least contributing to the observed host bias.

%------------------------------------
% color vs. metallicity as culprit
\subsection{Dependence of \snia\ Color on Progenitor Metallicity}
\label{sec:color_metal_trend}
\snf\ data show a strong correlation of \snia\ light curve color (SALT2 $c$) with host galaxy gas-phase metallicity, stronger than the analogous trend with host mass. This indicates a possible trend of \snia\ intrinsic color with metallicity which could easily drive some of the observed \snia\ Hubble residual bias.

\begin{figure}
\begin{center}
\includegraphics[width=0.45\textwidth]{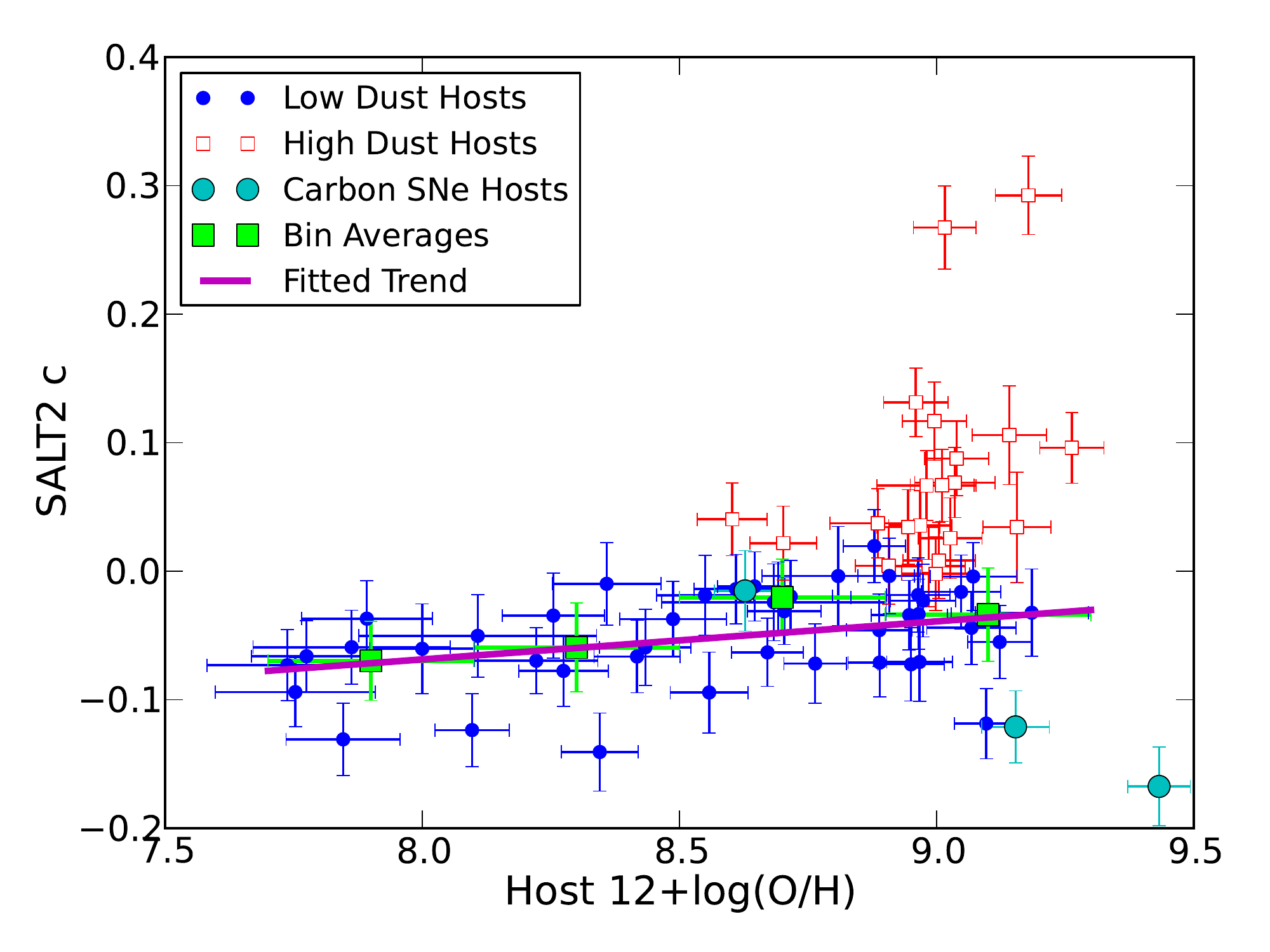}
\end{center}
\caption{SALT2 color $c$ vs. host galaxy metallicity for \snf\ \snia\ hosts. In bins of metallicity 0.4~dex wide, we fit for the weighted mean color (large green squares) of \snia\ hosts (small filled blue circles) after the rejection of extreme red outliers (small open red squares). The best fit trend fitted from these binned averages is shown as the thick magenta line.}
\label{fig:salt2_color_vs_metal}
\end{figure}

We quantify this color-metallicity trend here as follows. Because dust reddening in galaxies scales with mass and thus metallicity, we first isolate the subset of \sneia\ with extremely low reddening. We binned the data in bins of host metallicity of width 0.4~dex in the range $7.7 \leq 12+\log(O/H) \leq 9.3$, then used the aforementioned dust models (see Section~\ref{sec:two_color_effects}) to calculate what fraction of \sneia\ in that bin would be reddened by more than 0.05~mag. The lowest two metallicity bins had no outliers rejected (less than 5\% were expected to have reddening above our cut) while the highest metallicity bin had 65\% of its \sneia\ excised. In Figure~\ref{fig:salt2_color_vs_metal}, we plot the \snia\ color (SALT2 $c$) versus host metallicity, with those hosts passing the outlier rejection shown as small filled blue circles and those which failed this cut as open red squares.

We note here that three \sneia\ in our sample were identified as showing signatures of unburned carbon \citep{rcthomas11}, and these SNe are designated by large cyan circles (note none of these failed the outlier cut). \citet{rcthomas11} noted that these \sneia\ showed exceptionally blue colors, and indeed two of them appear significantly bluer than the color-metallicity locus. Since they do not significantly pull the trend we wish to fit, they are left in our sample, but we note that this subclass of \sneia\ may not obey the apparent color-metallicity relation we have identified here in our low-extinction subsample.

We use the sample of \sneia\ deemed to be ``low-reddening'' as described above to derive a linear fit to the trend of \snia\ color versus host metallicity, which is shown as the thick magenta line in Figure~\ref{fig:salt2_color_vs_metal}. The slope of the fitted trend is $0.028\pm0.010$ mag/dex, such that an \snia\ with 1 dex higher metallicity would be 0.028 magnitudes redder (in SALT2 $c$, which scales directly with observed \snia\ $B-V$ colors). 

% literature discussion!
% BVR to get SALT2 c
This trend of \snia\ intrinsic color with metallicity may lend some credence to theoretical explorations of the effect of metallicity on \snia\ color. Our measurement of $0.028\pm0.010$ mag/dex compares well to the value of 0.025 mag/dex given by the models of \citet{dominguez01} over a comparable range in metallicity. \citet{hoflich98} and \citet{sauer08} found that an increase in the progenitor metallicity for an \snia\ tended to yield redder colors in the optical, consistent with the trend we observe here, though \citet{lentz00} found a contradictory trend in their models. We note that our light curve colors were derived from fits to rest frame $BVR$ photometry, so our colors are not sensitive to the UV which may be more strongly affected by metallicity than the optical \citep{hoflich98, lentz00, sauer08}.
% Foley results
Interestingly, \citet{foley12} recently reported a correlation of the calcium velocity with host galaxy mass for a sample of \sneia. If their previously stated assumption that the calcium velocity correlates directly with intrinsic \snia\ color \citep{fk11} holds true, then this would agree qualitatively with our color-metallicity trend measured from \snf\ data.

% return to HR implications
Let us revisit the step in Hubble residuals found when splitting the sample into low and high metallicity hosts at $12+\log(\mathrm{O/H})=8.8$. The mean metallicity of the high metallicity hosts is $\left<12+\log(\mathrm{O/H})\right>_{high} = 9.02$, and the mean of the low metallicity hosts is $\left<12+\log(\mathrm{O/H})\right>_{low} = 8.35$. Given the trend just fitted, this implies that the intrinsic colors of high-metallicity-hosted \sneia\ are on average 0.019~mag redder than those in low-metallicity-hosted \sneia\ (for the distribution of metallicities in our sample). 

This possible intrinsic color discrepancy between \sneia\ in hosts of different metallicities may provide an explanation for the observed bias of corrected \snia\ luminosities with host mass (and metallicity). Let us consider the simplest scenario where the stretch-corrected peak luminosity of \sneia\ is independent of metallicity, but their intrinsic colors depend on metallicity. \snia\ cosmology analyses typically use a color-luminosity correction factor of about $\beta\sim3$, which implies that the luminosities of \sneia\ in high-metallicity hosts may be over-corrected for their color by about $\Delta m = 0.057$~mag. The fact that the observed Hubble residual step between high- and low-metallicity hosted \sneia\ is larger than this could indicate that other effects may be contributing to the observed \snia\ host bias.

The full picture of \snia\ intrinsic color and its implications for color correction techniques is likely to be more complicated, including such effects as were outlined in Section~\ref{sec:two_color_effects}. Our main interest is the structure of \snia\ Hubble residuals as a function of host mass produced by this trend, which we show in Figure~\ref{fig:dmu_trend_with_models} (dashed green curve). This model does not accurately trace the full structure of the observed host bias, but interestingly this model predicts a flattening of the Hubble residual trend at high host mass ($\log(M_*/M_\odot)\sim 11.0$) similar to that seen in the data. This is due to the flattening of the galaxy mass-metallicity relation at this mass scale, and may point to the importance of metallicity in the structure of the \snia\ Hubble residual host trend.

%------------------------------------
% prompt / tardy turnover
\subsection{Young versus Old \snia\ Progenitor Ages}
\label{sec:age_based_host_bias}
While the interplay of \snia\ intrinsic colors and dust are intriguing candidates for explaining the observed host bias, another potential source of the bias may arise from \snia\ progenitor ages. Studies have shown that the \snia\ rate in a given galaxy depends on both its stellar mass and star-formation rate, and we show here that the way those quantities are distributed along the galaxy mass sequence has direct implications for \snia\ progenitor ages.

The distribution of stellar mass along the galaxy mass sequence is well-described by a (modified) \citet{schechter76} function, and was first compared to the \snia\ host luminosity function in \citet{gallagher05}. Less massive galaxies are more intensely star-forming than their high-mass counterparts \citep{salim07, elbaz07}, making the distribution of star-formation along the galaxy mass sequence peak at lower masses and have a stronger low-mass tail.
Following the \snia\ rate dependence on these two quantities, the \snia\ host mass distribution can be fit as a linear combination of the stellar mass \citep{bell03, borch06} and star-formation \citep{salim07, elbaz07} distributions as measured from SDSS.

\begin{figure}
\begin{center}
\includegraphics[width=0.45\textwidth]{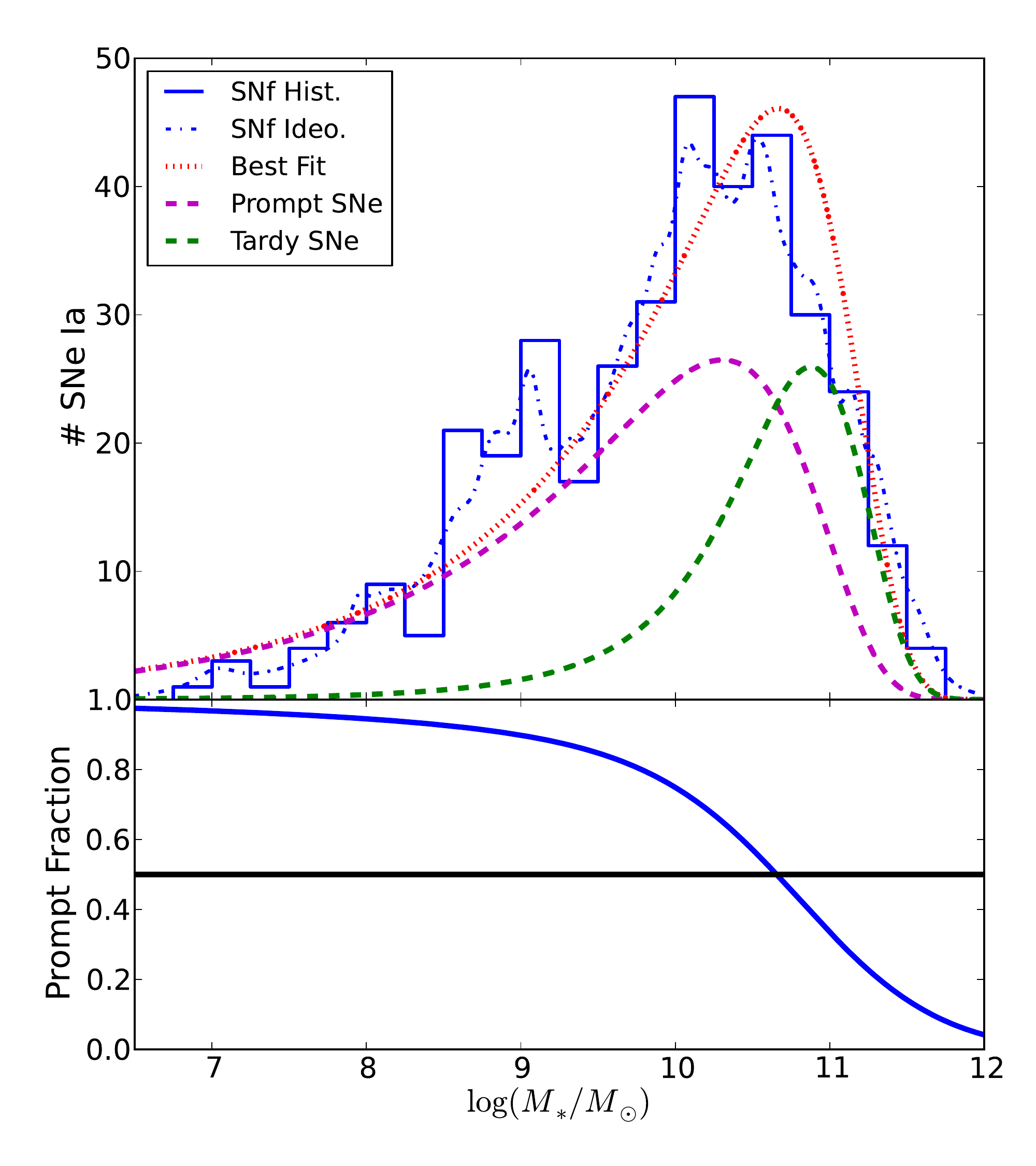}
\end{center}
\caption{Top: Histogram (solid blue line) and generalized histogram (dashed blue line) of \snia\ host galaxy masses from the full sample of \sneia\ discovered by the \snf\ search, compared to a fit (dashed red curve) composed of a component proportional the distribution of stellar mass (dashed green curve) in the local universe \citep{schechter76, bell03, borch06} and a component proportional to the distribution of star formation (dashed magenta curve) in the local universe \citep{salim07, elbaz07}. Bottom: Ratio of young \snia\ progenitors (magenta curve in top panel) to all \sneia\ (red curve in top panel) as a function of mass. \sneia\ undergo a transition to older progenitors at about $10^{10.5}M_\odot$.}
\label{fig:mass_dist_fit}
\end{figure}

% show plot of prompt fraction vs. mass!
We show in the top panel of Figure~\ref{fig:mass_dist_fit} the results of fitting the host mass distribution for the entire sample of \sneia\ discovered by \snf\ with these two components. We note here that the only free parameter in this fit is the relative ratio of the two components, as their shapes and positions along the galaxy mass scale are fixed by measurements of large external galaxy samples. This fit yielded a measurement of the fraction of \sneia\ arising from young progenitor systems (i.e. the ``prompt'' fraction) of \snfpromptfrac. This value was consistent with that derived from \snia\ rate coefficients from other studies \citep{scan05, smith12}. 

In the lower panel of the same Figure we show the prompt fraction as a function of host galaxy mass, which shows that the fraction of \sneia\ from young progenitors remains very high ($> 90$\%) until it undergoes a smooth transition to being subdominant over the mass range $10.0 \leq \log(M_*/M_\odot) \leq 11.0$. This shows that \sneia\ in low mass hosts originate primarily from young (``prompt'') progenitors while the majority of  \sneia\ in high mass hosts come from older (``tardy'') progenitors, with the equality point being reached at about $\log(M_*/M_\odot) = 10.5$. We note that this coarse two-component model would be better investigated with a full \snia\ delay time distribution and its implication for \snia\ ages along the galaxy mass sequence, and will be presented in the Paper~IV.

% comment on transition rapidity compared to data
Thus we see that \sneia\ undergo a significant change in progenitor age at a galaxy mass scale that is close to scale at which corrected \snia\ Hubble residuals undergo a transition. This could be a potential explanation of the observed host bias in \snia\ Hubble residuals if the luminosity of \sneia\ from young progenitors is different from that of old progenitors, or alternatively if the mean colors of these two subsets of \sneia\ differ. This could be a dependence of \snia\ luminosity (or color) on age, or it could be indicative of two different \snia\ progenitor channels which operate on different timescales and have different mean luminosities.

If the prompt/tardy transition is the driver of the host bias, we might expect the shape of the prompt fraction transition to reflect that seen in the Hubble residual data. We thus fit the prompt fraction model to the observed Hubble residual data by fitting for a mean prompt Hubble residual value and a mean tardy Hubble residual value. Minimizing the $\chi^2$ of this model yields a best fit luminosity difference between the prompt and tardy \sneia\ of 0.15~mag. This best fit model is shown in Figure~\ref{fig:dmu_trend_with_models}. The steepness of the prompt fraction transition appears slightly shallower than the observed structure in the \snia\ Hubble residual trend with host mass, so this model may be too simple. However, this trend is closer to the observed trend than the other models we considered, so we examine some important implications if it is true.

% field galaxy age transitions
This transition host mass for \snia\ progenitor ages is perhaps unsurprising given that the normal galaxy population undergoes a transition from primarily active to passive galaxies at this same mass scale. From SDSS data, \citet{bell03} and \citet{baldry04} fit the stellar mass functions  -- which are well fit by classical \citet{schechter76} functions -- for red (passive) and blue (active) galaxies. Both studies found that the number density (and thus mass density) of red and blue galaxies become equal at a galaxy mass scale of about $\log(M_*/M_\odot) = 10.3$. A similar transition between red and blue galaxies is also evident at higher redshifts \citep{borch06, bundy06, brammer11}, with the transition mass increasing as redshift increases.

If the \snia\ host bias is indeed driven by progenitor age, then the evolution in redshift of this galaxy color (and thus age) transition mass has important consequences for \snia\ luminosity corrections. For example, if \snia\ host masses are used to correct luminosities via a stepwise correction where luminosities of \sneia\ in hosts below some specific mass are adjusted with a constant shift, then the evolution of the galaxy age transition mass might imply that the \snia\ threshold host mass must necessarily evolve in redshift to avoid bias.

If progenitor age is indeed the cause of the observed host bias, then failure to correct \snia\ luminosities will result in cosmological biases due to the evolution of mean \snia\ progenitor ages. Suppose for simplicity that the luminosity offset between high- and low-mass hosted \sneia\ implies that young (prompt) \sneia\ are intrinsically fainter than old (tardy) \sneia, then the shift in the relative fraction of these two subsamples between high and low redshift will mean that the average corrected luminosities of \sneia\ differ at high and low redshift. Using the mean mass-SFR relation at high-redshift measured by \citet{elbaz07} adjusted to the same IMF as ours, we calculate that by $z\sim 1$ the fraction of prompt \sneia\ should increase to 95\% \citep[this is similar to the result of][ who noted that the fraction of low stretch, i.e. ``tardy'', \sneia\ should decrease from about 25\% to about 1\% beyond redshift $z\sim1$]{howell07}. Thus if the two sub-populations of \sneia\ differ in luminosity by 0.15, then an increase of 25\% of the prompt fraction implies \sneia\ at high redshift will be about 0.038~mag fainter than those at low redshift after stretch and color corrections.

%------------------------------------
\subsection{Metallicity Dependence of \snia\ \nifs\ Yield and Rise Time from \citet{bravo10}}
\label{sec:bravo_models}
The correlation of galaxy mass with metallicity has made metallicity a leading contender for explaining the bias of \snia\ Hubble residuals with host mass. In \snia\ models, metallicity affects the opacity and thus the resulting colors \citep{dominguez01,kasen09}, and the dynamics leading to explosion may also be affected \citep{townsley09}. The largest effect may arise from the predicted decrease in the production of \nifs\ due to an increase in neutronization with metallicity \citep{tbt03}. In related work, \citet{piro08} have suggested that additional neutronization occurs during the ``simmering'' phase, such that metallicity-induced neutronization will not become apparent until metallicities above $2/3 Z_\odot$ are reached.

Rather than attempt to transform our data into model space, we employ the results of \citet{bravo10} to determine relative brightnesses after stretch correction as a function of metallicity predicted by theoretical calculations.  \citet{bravo10} provide formulae for predicted peak brightness and lightcurve width, allowing a stretch correction to be applied to the models, as with the observations.  \citet{bravo10} account for burning that does not reach nuclear statistical equilibrium and find that neutronization during the simmering phase does not strongly affect the relative brightness trend with metallicity. They calculate models with and without a metallicity dependence for the deflagration to detonation transition density $\rho_{DDT}$; the former model produces a stronger dependence on metallicity.

\begin{figure}
\begin{center}
\includegraphics[width=0.45\textwidth]{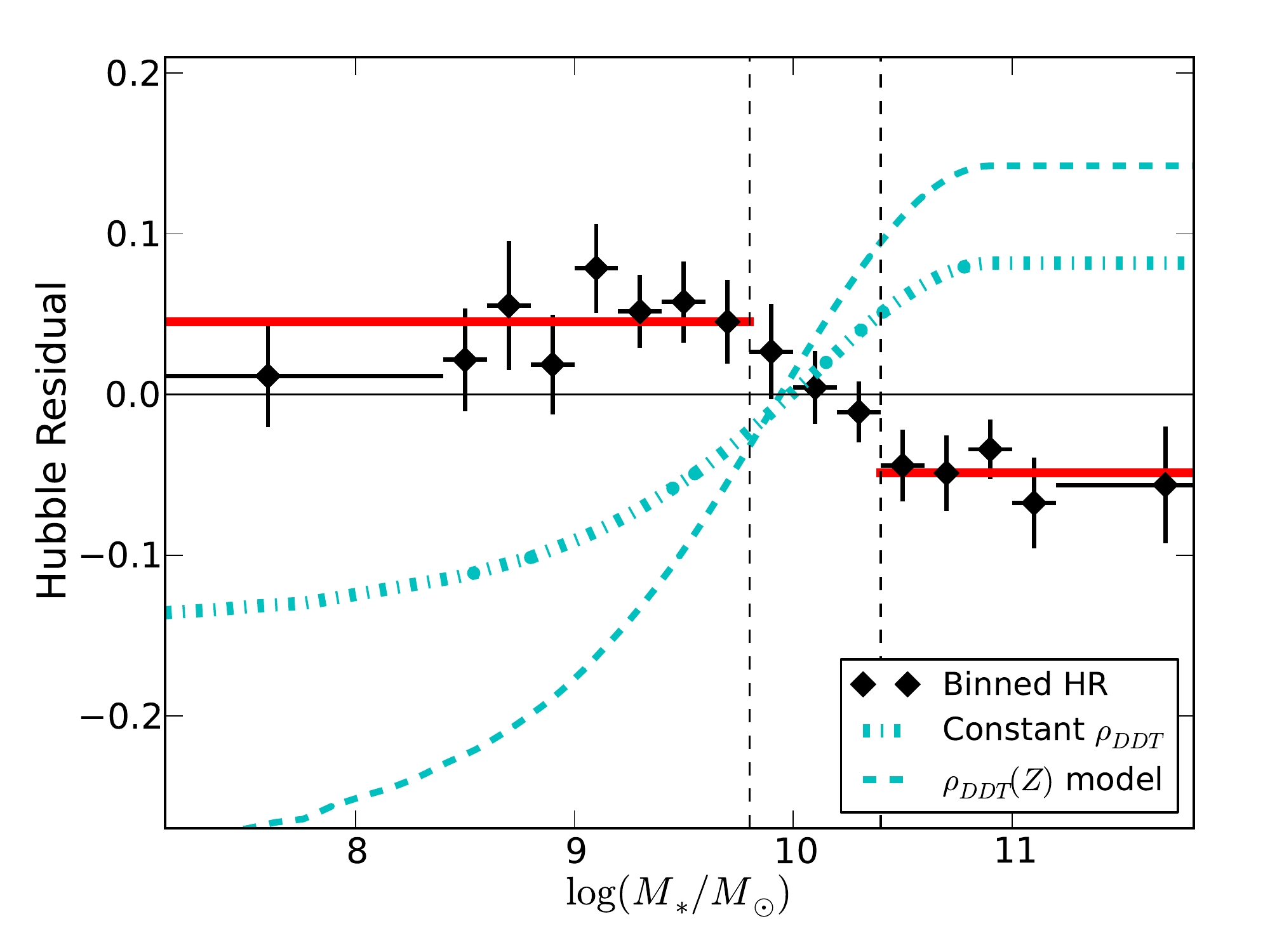}
\end{center}
\caption{Binned average Hubble residuals, as in Figure~\ref{fig:dmu_mass_fine_bins}, compared to the two models of \citet{bravo10}: the constant $\rho_{DDT}$ model (dashed-dotted cyan curve) and the metallicity dependent $\rho_{DDT}$ model (thin solid cyan curve). For reference we also show the Hubble residual averages for the two ``plateau'' regions as the thick red lines.}
\label{fig:bravo_models}
\end{figure}

In Figure~\ref{fig:bravo_models} we plot both \citet{bravo10} models (note the expanded y-axis scale). While the \citet{bravo10} models make predictions for bolometric quantities, we have found with \snf\ data (Scalzo et al., 2013, in prep.) a linear mapping between the $B$-band and bolometric quantities. This implies that the $B$-band and bolometric model predictions should be similar and it is worthwhile to compare the \citet{bravo10} predictions directly with the $B$-band Hubble residuals. It is quickly apparent that both models predict a trend of corrected luminosity with host mass that is {\em opposite} to that observed in the data.  This is due to the dimming of \sneia\ from the neutronization effect at high metallicity having a larger effect on the final \snia\ luminosity than stretch.

Interestingly, \citet{kasen09} predict an effect whose sign is opposite to that predicted by \citet{bravo10} and coincides with the sign of bias seen in the data, as was previously noted by \citet{sullivan10}. In those more detailed models the metallicity has a stronger effect on the lightcurve stretch than on the luminosity via the effect of temperature on opacity. Due to the computational demands of these models the sampling in metallicity is too coarse to make a quantitative comparison with our data at this time.

%------------------------------------
\subsection{Summary and Comparison of Models}
\label{sec:models_summary}
\begin{figure*}
\begin{center}
\includegraphics[width=0.85\textwidth]{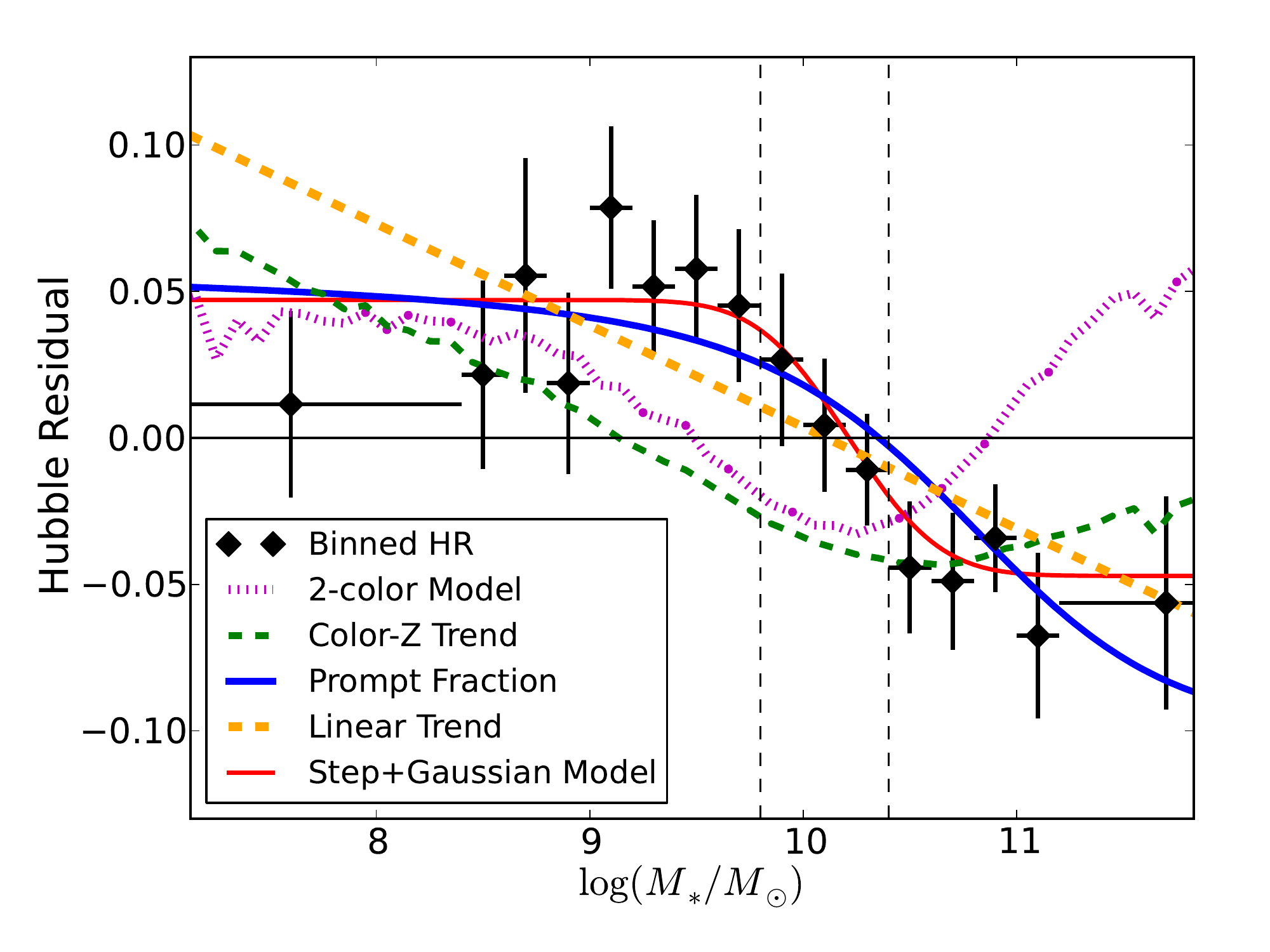}
\end{center}
\caption{Binned average Hubble residuals, as in Figure~\ref{fig:dmu_mass_fine_bins}, compared to predictions for several models: (i) simple two-color model of Section~\ref{sec:two_color_effects} (dotted magenta curve); (ii) trend of \snia\ intrinsic color with metallicity as discussed in Section~\ref{sec:color_metal_trend} (dashed green curve);  (iii) model where \sneia\ originate from ``prompt'' and ``tardy'' populations whose luminosities differ by the observed Hubble residual plateau offset, as described in Section~\ref{sec:age_based_host_bias}, and the ``prompt fraction'' of \sneia\ changes along the galaxy mass sequence; (iv) a simple linear trend fitted to the data (thick orange dashed line); and (v) low-mass and high-mass Hubble residual plateaus connected by an error function, which is exactly a step function convolved with a Gaussian (thin solid red lines).}
\label{fig:dmu_trend_with_models}
\end{figure*}

% alternative explanations - metallicity dependence of phillips relation?
We summarize the potential drivers of \snia\ host bias presented here. In Figure~\ref{fig:dmu_trend_with_models} we show the predicted shape of the mean \snia\ Hubble residuals versus host galaxy mass for the models presented above, as well as the best fit linear trend from Section~\ref{sec:combined_data_sets}. 

For the sake of comparison, we show a model represented by the two plateau values (i.e. constant Hubble residual value) outside the transition mass region of $9.8 \leq \log(M_*/M_\odot) \leq 10.4$ and connected with an error function. This functional form is exactly what would be expected for a step function convolved with a Gaussian, and our best fit for this functional form yielded a step function which transitions at $\log(M_*/M_\odot)=10.21$ convolved with a Gaussian of width $\sigma=0.47$~dex. This model, shown in Figure~\ref{fig:dmu_trend_with_models} as the thin solid red line, provides the best fit to the data. We note that the Gaussian width of $\sigma=0.47$~dex for this model is much larger than the typical measurement error of 0.1-0.2~dex for host masses, and indeed exceeds even the most conservative estimates of systematic stellar mass uncertainties (see Section~\ref{sec:host_corrections}). The driver of this transition width may indicate an astrophysical effect such as the width of the galaxy mass-metallicity relation (in mass) or the characteristic width of the blue-red galaxy transition (e.g. Section~\ref{sec:age_based_host_bias}). Greater statistics of \snia\ host galaxy metallicities or ages will likely be necessary to show whether the Hubble residual transition is sharper with respect to either of these properties.

% model comparison via chi2!!
In Table~\ref{tab:trend_models} we present the \chisq\ of the six models plotted in Figure~\ref{fig:dmu_trend_with_models} when compared to the full sample of 601 \sneia\ from the four data sets, as well as the improvement in \chisq\ for each model and the number of degrees of freedom $n_{dof}$.  Note that five cosmological parameters were applied to the data so that reduces the starting $n_{dof}$ to 596. For reference, we also report the \chisq\ if no trend is considered, which is exactly the \chisq\ of the standard cosmological fit for these SNe. 

% TABLE 3
\begin{table}
\begin{center}
\caption{\snia\ Host Bias Models vs. Data}
\label{tab:trend_models}
\begin{tabular}{lrrr}
\hline
Model               & $\chi^2$ & $\Delta\chi^2$ & $n_{dof}$ \\
\hline
No trend            &  $628.6$ & $  0.0$ & $596$ \\
2-Color Model       &  $637.1$ & $ +8.5$ & $596$ \\
Color-Z Trend       &  $629.8$ & $ +1.2$ & $594$ \\
Prompt / Tardy      &  $601.5$ & $-27.1$ & $594$ \\
Linear Model        &  $608.3$ & $-21.0$ & $594$ \\
Step+Gaussian Model &  $597.4$ & $-31.2$ & $594$ \\
\hline
\end{tabular}
\end{center}
\end{table}

% CONCLUSIONS OF WHICH MODELS ARE BEST!!!
Our first conclusion is that the 2-color and color-metallicity models are unlikely to be the sole driver of the observed host bias. However, the trend of \snia\ reddening by foreground host galaxy dust is based on observationally determined galaxy properties \citep{lee09, garn10}, and \snia\ colors are probably being affected by this trend to some degree. If so, the magnitude of the effect depends on the ability to successfully recover the dust reddening for each SN, which in turns depends on knowledge of the intrinsic colors of \sneia\ and the true dust extinction law. The color-metallicity trend agrees well with theoretical models and some observations (see Section~\ref{sec:color_metal_trend}) so may still be affecting \snia\ color corrections.

Our second conclusion from this analysis is that the \snia\ progenitor age model presents the best quantitative agreement with the observed host bias of all the physical models. This underscores the importance of considering age as a contender for driving the host bias. Thirdly we found that the \citet{bravo10} neutronization model exhibits poor agreement with the \snia\ host bias structure, as it predicts a trend opposite to that observed in the data. Finally we note that the analytical error function model yielded the best fit to the data. Along with the favorable agreement of the prompt/tardy model, this result may indicate that the \snia\ Hubble residuals are better represented by a bimodality with respect to their host properties rather than a continuous linear trend.

% other explanations?
We note that other scenarios may exist that could provide an explanation for the observed \snia\ host bias. For example, if the absorption properties of dust (i.e. $R_V$) vary with metallicity (a troubling possibility for standard candle cosmology), this could provide a less orthodox explanation of the observed trend of \snia\ $\beta$ with host mass, and would certainly result in biased \snia\ color corrections \citep[but see][]{chotard11}. We have thus far neglected possible effects originating from reddening of \sneia\ by some circum-stellar medium (CSM) deposited in the immediate SN vicinity by material shed in the later stages of the progenitor system evolution \citep{goobar08}. If the CSM has a different $R_V$ than the ISM dust, especially if CSM dust content depends on progenitor metallicity, then this could produce similar color correction errors as those predicted in our two color model of Section~\ref{sec:two_color_effects}. These and other explanations are worth investigation, but detailed discussions of their implications are beyond the scope of this work.

% SUMMARY
In summary, we demonstrated that dust, age, and metallicity all vary along the galaxy mass sequence. While dust effects alone could not quantitatively explain the observed \snia\ host bias, dust is probably still affecting \snia\ luminosity corrections in a small but important way. Metallicity-based effects show a characteristic flattening of Hubble residuals at high galaxy masses, due to the flattening of the galaxy mass-metallicity relation, which matches the observed \snia\ host bias structure. This may point to metallicity as a leading contender for driving the host bias. However we also showed that the likely evolution of \snia\ progenitor age along the galaxy mass sequence also shows favorable agreement with the data, so age should not be ruled out as a possible driver of the host bias.

None of these effects acting alone seem able to mimic a transition as steep as that observed in the binned \snia\ Hubble residuals. It may be the case that the age, metallicity, and dust trends are all present in some combination. We believe that discrimination between these properties will require either (or probably both) a more detailed study of the immediate \snia\ environments or a complete accounting of intrinsic \snia\ color variation via constraints on host galaxy dust reddening. A larger sample of measured \snia\ host ages and metallicities may elucidate whether the Hubble residual host bias is more strongly correlated with these properties than with host mass.

%%%%%%%%%%%%%%%%%%%%%%%%%%%%%%%%%%%%%%%%%%%%%%%%%%%%%%%%%%%%%%%%%%%%%%%%%%%%%
\section{Correcting Host Bias in \snia\ Data}
\label{sec:host_corrections}
In light of the observed \snia\ luminosity host bias, a major concern for future \snia\ cosmological studies is how to effectively correct for this effect in \snia\ data. Currently, the observed correlation of \snia\ Hubble residuals with their host galaxy masses has motivated the possible use of \snia\ host mass as a third brightness correction parameter (in addition to stretch and color) in \snia\ cosmological analyses. Indeed, some authors have already performed \snia\ cosmology fits using a linear host-based brightness correction \citep{sullivan11a} or a step-wise brightness correction for host mass where brightnesses of low-mass-hosted \sneia\ are adjusted by a constant amount \citep{suzuki12}.

% summary
In this Section we briefly describe the potential contribution of host corrections to the \snia\ cosmology error budget, as well as potential sources of bias in \snia\ cosmology if host corrections are not used. In particular, we seek to estimate two quantities: (i) the amount of error introduced to \snia\ Hubble residuals from the host correction itself, and (ii) the bias in mean luminosity of high-redshift ($z\sim1$) \sneia\ if host corrections are not applied.

Metallicity is currently perceived to be a leading candidate for explaining the origin of the \snia\ host bias, but we showed above that dust and progenitor age may be driving part or all of the observed host bias. We will focus on dust effects in Paper~III, and age effects in Paper~IV, but here we explore the potential implications if the trend is metallicity-driven. 

Specifically, we consider a model where metallicity affects the luminosity of \sneia\ in a linear way with a slope equal to that measured from \snf\ data in Section~\ref{sec:data}, and assume that \snia\ luminosity corrections will be derived from photometrically-estimated host galaxy stellar masses. The major sources of systematic errors under this scheme would be the systematics in galaxy stellar population synthesis (SPS) techniques to estimate host masses, the dispersion in (and evolution of) the galaxy MZ relation, and the difference between the inferred average metallicity of the \snia\ host galaxy and the metallicity of the SN progenitor itself. We address and quantify each of these effects in turn.

%--------------------
% error from SPS systematics
\subsection{SPS systematics}
If host galaxy properties are used to correct \snia\ luminosities, then the \snia\ error budget will be increased by the systematics involved in this process. In particular, if host masses are used as correction factors, then the systematics accompanying the application of stellar population synthesis (SPS) techniques will impact \snia\ cosmology. Though a census of the full body of SPS systematics is beyond the scope of this work \citep[see thorough discussions in, e.g.,][regarding mass-to-light ratio recovery efficiencies and the impact of stellar models on fitted mass-to-light ratios]{gb09, cgw09}, we summarize here a few key elements that will surely impact host-based \snia\ luminosity corrections.

% tp-agb etc. from conroy = 0.3 dex
\citet{cgw09} showed that uncertainties in the contribution of TP-AGB stars to galaxy luminosities and uncertainties in the shape of the stellar initial mass function (IMF) can result in stellar mass uncertainties of order 0.3~dex.
% recovery efficiency from my work = 0.2 dex
The ability to recover mass-to-light ratios is limited by how well photometry can constrain the galaxy star formation history (SFH). We showed in Paper~I that full photometric coverage from UV to the NIR can provide mass-to-light ratios with a dispersion of about 0.1~dex, but even a single color constrains these to about 0.3~dex.
% assumptions (i.e. prior) on SFHs
\citet{gb09} showed that erroneous assumptions about priors on galaxy SFHs, such as the frequency and intensity of bursts of star formation, can bias mass-to-light ratio values by as much as about 0.2~dex.

% summary
Considering these effects together, galaxy SPS techniques carry with them a systematics floor that may be as high as 0.4~dex in stellar mass. We note that this is a conservatively high value for SPS uncertainties if one assumes that all galaxy masses are affected by TP-AGB, SFH, and IMF uncertainties at the maximum level estimated by previous authors. If the stellar mass uncertainty were truly this high, then it would encompass almost the entirety of the 0.47~dex width of the step function Hubble residual model from Section~\ref{sec:models_summary}. This would require a physical model of \snia\ Hubble residuals that is a true step function in host \emph{mass}, which is very unlikely. Typical host mass uncertainties for the \snf\ sample are at the 0.15~dex level and are driven by SFH uncertainty (Section~\ref{sec:data}, also Paper~I), but we calculate this maximal host mass error of 0.40~dex to demonstrate the maximal cosmological uncertainty host mass estimates might introduce. 

Propagating this uncertainty in the host mass estimate through our fitted trend of \snia\ luminosities with mass implies that the resultant uncertainty in the host-correction terms for \snia\ luminosities will have an additional scatter of about 0.013~mag. We note than any coherent error in the estimation of host galaxy stellar masses would be absorbed into the standardized \snia\ luminosity (i.e. $M_B$) so long as it is applied consistently across all redshifts, and thus would not bias the measurement of cosmological parameters.

%--------------------
% error from dispersion in (and evolution of) the MZ relation
\subsection{Dispersion in (and evolution of) the MZ relation}
% MZ dispersion
The use of photometric \snia\ host galaxy masses as proxies to correct for a metallicity-driven effect will have some systematic errors resulting from the dispersion in the galaxy MZ relation. At high metallicities, this dispersion is as low as 0.1~dex, but increases to about 0.3~dex at low metallicities \citep{trem04}. We use the observed trend of \snia\ Hubble residuals with metallicity from \snf\ (Section~\ref{sec:salt_trends}) to calculate that, even at the extreme high MZ dispersion of 0.3~dex, the MZ relation dispersion would introduce a dispersion in host-mass-corrected \snia\ luminosities of about 0.032~mag.

% MZ EVOLUTION - if we use mass, what is Z bias vs. z from MZ evolution?
If indeed the observed host bias is driven by progenitor metallicity, then the use of \snia\ luminosity corrections derived from host galaxy masses must account for evolution of the MZ relation with redshift in order to avoid biasing measurements of cosmological parameters. \citet{zahid11} measured the galaxy MZ relation at $z\sim0.8$ from the DEEP2 survey and found that galaxies at high redshift have metallicities approximately 0.15~dex lower than their counterparts in the local universe. Again using the host bias trends measured in Section~\ref{sec:salt_trends}, this would imply a systematic bias of about 0.016~mag between the corrected luminosities of low- and high-redshift \sneia. Though very small, this bias would be present if \snia\ host masses are used to correct a metallicity-driven \snia\ luminosity effect without converting to the redshift-dependent metallicity implied by the evolving MZ relation.

% NO HOST CORRECTIONS - what is bias from global Z vs. z?
Finally we consider the potential bias in \snia\ luminosities if host-based corrections are \emph{not} applied. The main driver of this bias is the difference in average host galaxy mass between the low- and high-redshift \snia\ samples. For example, the mean host mass and redshift of the low redshift \snia\ sample from \citet{kelly10} are $z = 0.035$ and $\log(M_*/M_\odot) = 10.67$, while for the high-redshift SNLS sample of \citet{sullivan10} these are $z = 0.634$ and $\log(M_*/M_\odot) = 9.79$. Using the MZ relation of T04 and a MZ offset of 0.1~dex (scaled down approximately by redshift) due to its evolution with redshift, this implies that current samples of high-redshift \snia\ hosts are likely to be 0.29~dex more metal poor than the low-redshift sample. This would imply an offset in stretch- and color-corrected \snia\ luminosities of 0.031~mag in \emph{current} \snia\ samples.  However for \snf\ data at low redshift (with median stellar mass of $\log(M_*/M_\odot)=10.1$), this implies the offset between high and low redshift \snia\ host metallicities would be about 0.2~dex, corresponding to a magnitude offset of 0.021~mag.

%--------------------
% error from progenitor metallicity uncertainty
\subsection{Discrepancies between host and progenitor metallicities}
We now turn to the systematic error on host-corrected \snia\ luminosities introduced by the difference between the measured (or inferred) host metallicity and the metallicity of the \snia\ progenitor. 
% SOURCES OF Z DIFF
An individual galaxy has an internal distribution of stellar metallicities with non-negligible dispersion, often with gradients as high as 0.5~dex per scale radius \citep[e.g.][]{spolaor10}, and the SN progenitor metallicity is drawn at random from this distribution. Based on the metallicity trend observed above with \snf\ data (Section~\ref{sec:salt_trends}), a galaxy with internal metallicity dispersion of 0.5~dex would introduce a random error on \snia\ brightnesses of about 0.053~mag, which is much smaller than the observed dispersion in corrected \snia\ luminosities ($\sim0.15$~mag) or the offset between the corrected luminosities of high- and low-metallicity-hosted \sneia\ ($\sim0.1$~mag). Thus despite the potential differences between measured \snia\ host metallicities and the progenitor metallicities themselves, these would ultimately have only small effects on the corrected brightnesses of \sneia. The range of observed galaxy metallicities has a much larger span than the typical internal metallicity dispersion of a single galaxy, so ultimately such a correction would introduce an error whose magnitude is smaller than the bias it is intended to correct. 

It is also important to consider potential biases introduced by the difference between the \snia\ progenitor metallicity at the time of its formation and the metallicity of the host galaxy measured at the time of explosion. Chemical evolution in the host galaxy of a \snia\ means that the average current metallicity may be significantly higher than it was at the time the \snia\ progenitor system was formed (especially for long delay times of several Gyr), by as much as perhaps 0.6~dex \citep{bravo11}. If all \sneia\ have the same (average) offset between progenitor and host metallicity, then this offset could be absorbed into the standardized luminosity $M_B$. More specifically, so long as the average offset between host and progenitor metallicity remains constant at all redshifts, then the mean $M_B$ would be valid for all redshifts.

However, an evolution of the average SN-host metallicity offset would produce a bias between high- and low-redshift samples. This would most likely arise if the average delay time of \sneia\ evolves with redshift, which we expect it should. To explore this in a simple fashion, let us consider the simple segregation of \sneia\ into short delay times (``prompt'' \sneia) and long delay times (``tardy'' \sneia) and assume that all prompt \snia\ progenitors are at the same metallicity as their host whereas all tardy \snia\ progenitors are actually 0.6~dex lower metallicity than their hosts. In this scheme the average SN-host metallicity offset will change from low- to high-redshift proportionally to the fraction of tardy \sneia. As discussed above in Section~\ref{sec:age_based_host_bias}, we calculated in Childress et al. (2012c, in prep.) that the current prompt fraction at low redshift is about 70\%, while at high redshift it should increase to 95\% (due to the increasing intensity of star-formation). Thus the tardy fraction decreases by 25\%, so the average offset between SN and host metallicity would decrease by 0.15~dex. This means the average \snia\ luminosity zeropoint at high redshift would be brighter by about 0.016~mag.

%--------------------
% error from progenitor metallicity uncertainty
\subsection{Summary and Outlook}
In this Section we investigated possible contributions of host galaxy systematics to the scatter in \snia\ Hubble residuals if hosts are used as a third luminosity correction parameter, as well as the bias in high-redshift \snia\ luminosities if a host correction is {\em not} applied. We proceeded under the simple assumption that metallicity linearly affects the \snia\ luminosity, and calculated bias and dispersion values which are summarized in Table~\ref{tab:host_corr_errs}.

% TABLE 4
\begin{table}
\begin{center}
\caption{\snia\ Host Correction Errors}
\label{tab:host_corr_errs}
\begin{tabular}{lrr}
\hline
Effect                    & Non-Correction & Correction       \\
                          & Bias (mag)     & Dispersion (mag) \\
\hline
SPS Systematics            &                & \\
 - SFH Uncertainty         &  --            & 0.007 \\
 - TP-AGB etc.             &  --            & 0.011 \\
 - {\bf Total}             &  --            & 0.013 \\
\hline
MZ Relation                &                & \\
 - Mass-based Z-correction & --             & 0.032 \\
 - MZ Evolution (vs. z)    & 0.016          & -- \\
\hline
SN-Host Z Mismatch         &                & \\
 - Host Z dispersion       &  --            & 0.053 \\
 - Delay time Z evolution  & 0.016          & --    \\
\hline
\end{tabular}
\end{center}
\end{table}

The sources of dispersion in corrected \snia\ luminosities generally arise from systematic uncertainties in the estimation of host galaxy properties and the differences between global host metallicity and the metallicity of the SN progenitor itself. Even in the most pessimistic scenario, these effects have a quadrature sum of 0.063~mag, which is significantly smaller than the observed scatter in \snia\ Hubble residuals. Conversely, the failure to apply host corrections to \snia\ luminosities will bias \snia\ Hubble residuals at high redshift due to the evolution of mean progenitor metallicity at the level of 0.031~mag. We thus claim that host corrections should be applied for current \snia\ samples as these will correct a critical source of bias while contributing negligibly to the \snia\ Hubble residual scatter.

Finally, we caution that a proper accounting of the evolutionary effects of galaxy metallicity with redshift is critical to avoiding bias in \snia\ host corrections. Failure to incorporate a redshift-dependent galaxy mass-metallicity relation would bias high redshift \snia\ Hubble residuals by 0.016~mag. A similar amount of bias could be introduced if the mean delay time between the \snia\ progenitor production and its explosion causes a systematic offset between host and SN metallicity which is not accounted for. The former effect is well studied and can be directly compensated for, while the latter is dependent on the \snia\ delay time distribution which is currently still a topic of vigorous study.

%%%%%%%%%%%%%%%%%%%%%%%%%%%%%%%%%%%%%%%%%%%%%%%%%%%%%%%%%%%%%%%%%%%%%%%%%%%%%
\section{Conclusions}
\label{sec:conclusions}
Using \snia\ light curves and host galaxy properties from the Nearby Supernova Factory (\snf), we showed that the stretch- and color-corrected \snia\ luminosities exhibit a bias with respect to the masses, specific star-formation rates, and metallicities of their host galaxies. We also showed analogous trends exist for model-free observational quantities such as host absolute magnitude and color. These results are in good agreement with the findings of previous authors \citep{kelly10, lampeitl10, sullivan10, konishi11, gupta11, dandrea11}.

When we combine our results with those of \citet{kelly10}, \citet{sullivan10}, and \citet{gupta11}, we demonstrate that stretch- and color-corrected \snia\ Hubble residuals of \sneia\ in high- and low-mass hosts differ by \allmassmagstep, a $5.6\sigma$ detection of \snia\ host bias. Furthermore, we show that when Hubble residuals for these data sets are derived with common \snia\ luminosity correction parameters ($\alpha$, $\beta$) the shape the Hubble residuals trend appears to plateau at low and high host masses, with a difference of \allwidemassmagstep\ that transitions quickly in the host mass range $9.8 \leq \log(M_*/M_\odot) \leq 10.4$ (see Section~\ref{sec:hr_trend_structure} and Figure~\ref{fig:dmu_mass_fine_bins}).

%------------------
% possible host bias origins
We then examined several physical effects which could potentially drive the observed trend of \snia\ Hubble residuals with host galaxy properties. Most importantly we highlighted the fact that dust, age, and metallicity all vary along the galaxy mass sequence and could contribute to the \snia\ host bias. Dust alone could not explain the entire bias observed, but both metallicity and age could produce Hubble residual trends with similar magnitude and structure to those observed in the data.

%------------------
% Host corrections in SN Ia cosmology
Finally we considered the cosmological implications of using (or not using) \snia\ host properties to correct their luminosities. We showed that systematic errors in the estimation of \snia\ host galaxy properties would yield a maximum total dispersion of 0.06~mag, but failure to correct the observed Hubble residual host trends would bias the luminosities of high-redshift ($z\sim1$) \sneia\ by about 0.02-0.04~mag. We thus concluded that host corrections in \snia\ cosmology contribute negligibly to the \snia\ error budget but correct a critical source of bias and are recommended to be used at this time. However we cautioned that a full understanding of the astrophysical origin of the bias is necessary in order to derive a correct redshift dependence of the host corrections.

%------------------
% synthesis - comment on possibility of multiple effects
% and make prescriptions for observational data to discriminate
% measure dust! intrinsic color!
Correction of the observed \snia\ Hubble residual host bias is a critical step for the future of \snia\ cosmology. The obvious current means to account for this bias is to apply a third empirical \snia\ luminosity correction derived from host galaxy properties, but the ideal future goal would be to uncover the physical origin of this trend and find a more direct means of correcting it. We have shown here that many aspects of \snia\ progenitors and environments change along the galaxy mass scale and may be altering the observed luminosities and colors of \sneia\ through several mechanisms. We suggest that the preferred next step for investigating the \snia\ host bias is to build a sample of \sneia\ whose environmental properties (particularly age, metallicity, and dust reddening) can be constrained on a local scale. The ideal hunting ground for such \sneia\ is in the low redshift universe where measurement of spatially resolved galaxy properties is possible.

%%%%%%%%%%%%%%%%%%%%%%%%%%%%%%%%%%%%%%%%%%%%%%%%%%%%%%%%%%%%%%%%%%%%%%%%%%%%%
% Facilities
{\it Facilities:} 
\facility{UH:2.2m},
\facility{GALEX},
\facility{Shane (Kast Double spectrograph)}, 
\facility{Keck:I (LRIS)},
\facility{Blanco},
\facility{SOAR},
\facility{Gemini:South}

%%%%%%%%%%%%%%%%%%%%%%%%%%%%%%%%%%%%%%%%%%%%%%%%%%%%%%%%%%%%%%%%%%%%%%%%%%%%%
\vskip11pt
% Acknowledgements
\scriptsize
% funding
Acknowledgments: Based in part on observations made with the NASA Galaxy Evolution Explorer. \galex\ is operated fro NASA by the California Institute of Technology under NASA contract NAS5-98034. The authors graciously acknowledge support from \galex\ Archival Research Grant \#08-GALEX508-0008 for program GI5-047 (PI: Aldering).
This work was supported by the Director, Office of Science, Office of High Energy Physics, of the U.S. Department of Energy under Contract No. DE-AC02-05CH11231; the U.S. Department of Energy Scientific Discovery through Advanced Computing (SciDAC) program under Contract No. DE-FG02-06ER06-04; by a grant from the Gordon \& Betty Moore Foundation; in France by support from CNRS/IN2P3, CNRS/INSU, and PNC; in France by support from CNRS/IN2P3, CNRS/INSU, PNC, and Lyon Institute of Origins under grant ANR-10-LABX-66; and in Germany by the DFG through TRR33 ``The Dark Universe''. 
This research used resources of the National Energy Research Scientific Computing Center, which is supported by the Director, Office of Science, Office of Advanced Scientific Computing Research, of the U.S. Department of Energy under Contract No. DE-AC02-05CH11231.  We thank them for a generous allocation of storage and computing time. HPWREN is funded by National Science Foundation Grant Number ANI-0087344, and the University of California, San Diego. The Centre for All-sky Astrophysics is an Australian Research Council Centre of Excellence, funded by grant CE110001020.

% observing
The authors would like to thank the excellent technical and scientific staff at the many observatories where data was taken for this paper: the University of Hawaii 2.2m telescope, Lick Observatory, Keck Observatory, the Blanco 4m telescope, the SOAR telescope, and Gemini South. Some data presented herein were obtained at the W. M. Keck Observatory, which is operated as a scientific partnership among the California Institute of Technology, the University of California, and the National Aeronautics and Space Administration; the Observatory was made possible by the generous financial support of the W. M. Keck Foundation. We wish to recognize and acknowledge the very significant cultural role and reverence that the summit of Mauna Kea has always had within the indigenous Hawaiian community, and we are extremely grateful for the opportunity to conduct observations from this mountain. 
We also thank Dan Birchall for assistance with SNIFS observations.
We are very grateful to David Rubin for providing SALT2.2 light curve fits to the CfA light curves in advance of the forthcoming Union3 analysis.
% josh and dan kasen
We also thank Josh Meyers and Dan Kasen for useful discussions.
% referee thanks go here:
We thank the anonymous referee who provided very helpful comments.

% SDSS boilerplate
Some of the data analyzed here were obtained from the Sloan Digital Sky Survey Eight Data Release (SDSS-III DR8). Funding for SDSS-III has been provided by the Alfred P. Sloan Foundation, the Participating Institutions, the National Science Foundation, and the U.S. Department of Energy Office of Science. The SDSS-III web site is http://www.sdss3.org/. SDSS-III is managed by the Astrophysical Research Consortium for the Participating Institutions of the SDSS-III Collaboration including the University of Arizona, the Brazilian Participation Group, Brookhaven National Laboratory, University of Cambridge, Carnegie Mellon University, University of Florida, the French Participation Group, the German Participation Group, Harvard University, the Instituto de Astrofisica de Canarias, the Michigan State/Notre Dame/JINA Participation Group, Johns Hopkins University, Lawrence Berkeley National Laboratory, Max Planck Institute for Astrophysics, New Mexico State University, New York University, Ohio State University, Pennsylvania State University, University of Portsmouth, Princeton University, the Spanish Participation Group, University of Tokyo, University of Utah, Vanderbilt University, University of Virginia, University of Washington, and Yale University.

%%%%%%%%%%%%%%%%%%%%%%%%%%%%%%%%%%%%%%%%%%%%%%%%%%%%%%%%%%%%%%%%%%%%%%%%%%%%%
% BIBLIOGRAPHY
\bibliographystyle{apj}
\bibliography{snf_hubble_hosts}

%%%%%%%%%%%%%%%%%%%%%%%%%%%%%%%%%%%%%%%%%%%%%%%%%%%%%%%%%%%%%%%%%%%%%%%%%%%%%
\end{document}